\documentclass[prd,twocolumn,nofootinbib,preprintnumbers,superscriptaddress,showpacs]{revtex4-1}

\usepackage{amsfonts}
\usepackage{amsmath}
\usepackage{amssymb}
\usepackage{bm}
\usepackage{dcolumn}
\usepackage{graphicx}
\usepackage{graphics}
\usepackage[latin1]{inputenc}
\usepackage{latexsym}
\usepackage{rotating}
\usepackage{url}
\usepackage{xspace}
\usepackage[usenames]{color}
\usepackage{mathrsfs}
\usepackage{hyperref}
\usepackage{epstopdf}
\usepackage{tensor}
\usepackage[margin=1.0in]{geometry}
\usepackage[scientific-notation=true]{siunitx}
\usepackage{placeins}

\newcommand{\be}{\begin{equation}}
\newcommand{\ee}{\end{equation}}
\newcommand{\bal}{\begin{align}}
\newcommand{\bsp}{\begin{split}}
\newcommand{\esp}{\end{split}}

\newcommand{\GR}{{\mbox{\tiny GR}}}
\newcommand{\Hz}{{\mbox{\tiny H}}}
\newcommand{\new}{{\mbox{\tiny new}}}
\newcommand{\old}{{\mbox{\tiny old}}}
\newcommand{\eff}{{\mbox{\tiny eff}}}
\newcommand{\nonlin}{{\mbox{\tiny nonlin}}}
\newcommand{\ISCO}{{\mbox{\tiny ISCO}}}
\newcommand{\LR}{{\mbox{\tiny LR}}}

\newcommand{\bw}{\begin{widetext}}
\newcommand{\ew}{\end{widetext}}

\newcommand{\nn}{\nonumber}

\newcommand{\pd}{\partial}
\newcommand{\cd}{\nabla}
\newcommand{\tn}{\tensor}

\newcommand{\ssqth}{\sin^2{\theta}}

\newcommand{\mcl}{\mathcal}

\newcommand{\mrm}{\mathrm}
\newcommand{\rarr}{\rightarrow}

\newcommand{\lb}{\left(}
\newcommand{\rb}{\right)}
\newcommand{\lcb}{\left\{}
\newcommand{\rcb}{\right\}}
\newcommand{\lsb}{\left[}
\newcommand{\rsb}{\right]}

\begin{document}
\title{Exact Black Hole Solutions in Modified Gravity Theories: \\ Spherical Symmetry Case}

\author{Andrew Sullivan}
\affiliation{Department of Physics, Montana State University, Bozeman, MT 59717, USA}
\author{Nicol\'as Yunes}
\affiliation{Department of Physics, Montana State University, Bozeman, MT 59717, USA}
\author{Thomas P.~Sotiriou}
\affiliation{School of Mathematical Sciences \& School of Physics and Astronomy, University of Nottingham,
University Park, Nottingham, NG7 2RD, United Kingdom}

\date{\today}

\begin{abstract} 

Detailed observations of phenomena involving black holes, be it via gravitational waves or more traditional electromagnetic means, can probe the strong field regime of the gravitational interaction. 
The prediction of features in such observations requires detailed knowledge of the black hole spacetime, both within and outside of General Relativity. 
We present here a new numerical code that can be used to obtain stationary solutions that describe black hole spacetimes in a wide class of modified theories of gravity.
The code makes use of a relaxed Newton-Raphson method to solve the discretized field equations with a Newton's polynomial finite difference scheme.
We test and validate this code by considering static and spherically symmetric black holes both in General Relativity, as well as in scalar-Gauss-Bonnet gravity with a linear (linear scalar-Gauss-Bonnet) and an exponential (Einstein-dilaton-Gauss-Bonnet) coupling.
As a by-product of the latter, we find that analytic solutions obtained in the small coupling approximation are in excellent agreement with our fully non-linear solutions when using a linear coupling. As expected, differences arise when using an exponential coupling.
We then use these numerical solutions to construct a fitted analytical model, which we then use to calculate physical observables such as the innermost stable circular orbit and photon sphere and compare them to the numerical results.
This code lays the foundation for more detailed calculations of black hole observables that can be compared with data in the future.  
\end{abstract}

\maketitle

\section{Introduction}
\label{sec:intro}

The recent discovery of gravitational waves~\cite{PhysRevLett.116.061102, PhysRevLett.116.241103, PhysRevLett.118.221101, 2041-8205-851-2-L35, PhysRevLett.119.141101, PhysRevLett.119.161101} has inaugurated a new era of multi-messenger astrophysics that opens up an entirely new avenue to test Einstein's theory of General Relativity (GR) in the extreme gravity regime~\cite{Yunes2013}. GR is thus far in excellent agreement with experiments and observations. However, this statement does rely on the assumption that dark matter or dark energy do not signal a deviation from GR. Moreover, GR has so far resisted quantisation attempts. These considerations have provided motivation for exploring modified theories of gravity. 

Unsurprisingly, modified theories typically increase the complexity of the field equations to such a degree that the calculation of observable predictions becomes incredibly difficult and sometimes even seemingly impossible if working analytically. 
Historically, experimental tests that probe the weak-field regime prompted the calculation of solutions and observables through perturbation theory~\cite{Will2014}, and more recently, this has been further justified through effective field theory arguments~\cite{PhysRevD.79.044033, Burgess2004, doi:10.1146/annurev.nucl.56.080805.140508}. A by-product of assuming that GR modifications are small relative to GR predictions is that analytical calculations become tractable again. But once one begins to probe the extreme gravity regime, perturbative techniques need not be well-justified, as they typically eliminate strong-field instabilities that could have observational consequences. The typical example of this is spontaneous scalarization, a process through which large modifications to GR arise in a class of scalar-tensor theories when considering non-linear solutions for neutron stars above a certain compactness~\cite{PhysRevD.58.042001, PhysRevD.90.124091} or black holes within a certain mass range~\cite{PhysRevLett.120.131104, PhysRevLett.120.131103}. 

The need for more precise solutions and observable predictions without the use of perturbation theory then becomes clear, and we here take steps in this direction. We present a numerical infrastructure to produce exact solutions that describe stationary black holes in a wide class of modified theories of gravity.  This paper is focuses on describing the numerical setup and applying it in the simplest case of spherically symmetric solutions that can act as a benchmark. By ``exact" we mean solutions obtained numerically and without the use of perturbative techniques, including small coupling expansions. The code is computationally efficient, converging to an answer of prescribed accuracy within only a few iterations, and thus allowing for calculations in laptop-class computers. We validate this infrastructure against known analytic solutions in General Relativity, as well as in a specific member of the class of modified theories to which this infrastructure is applicable. 

\begin{figure*}[htb]
\begin{center}
\resizebox{8cm}{!}{\include{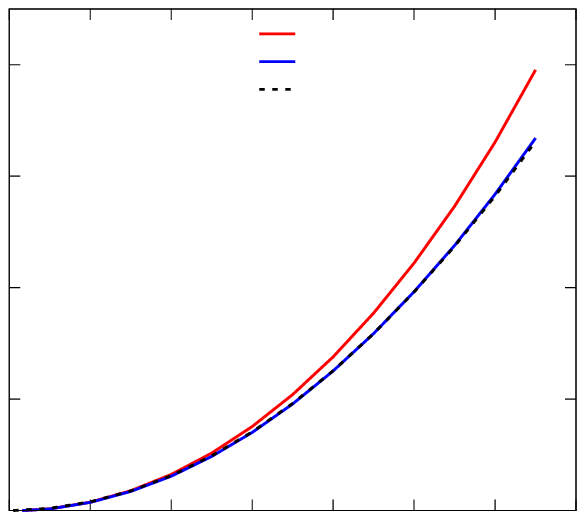}}
\resizebox{8cm}{!}{\include{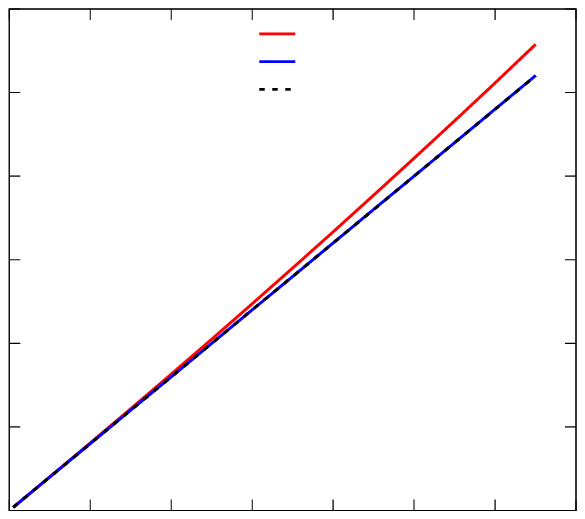}}
\caption{\label{fig:Masscomp} (Color Online) 
Relative fractional correction to the ADM mass (left) and the dimensionless scalar monopole charge as a function of the dimensionless sGB coupling parameter $\bar{\alpha}$. The solid red and blue lines correspond to our numerical solutions using an exponential coupling (EdGB) and a linear coupling (linear sGB) respectively, while the black dashed line corresponds to the analytical perturbative solution. Observe that the perturbative approximation agrees very well with both solutions for small $\bar{\alpha}$, and in fact, it agrees well with the numerical linear sGB solution for \emph{all} $\bar{\alpha}$ explored. However, the perturbative approximation differs from the numerical EdGB solution for $\bar{\alpha} \gtrsim 10^{-3}$, showing the breakage of the small-coupling approximation.}
\end{center}
\end{figure*}

The basic problem our numerical infrastructure solves is that of finding the solution to an elliptic system of nonlinear and coupled differential equations. The typical approach to solve this problem numerically is to discretize the differential equations through a finite difference scheme. This leads to a residual error on the solution of the system of equations that one wishes to minimize. Our algorithm employs a Newton polynomial method to discretize the equations, which appropriately allows for an easier analytical evaluation of the Jacobian of the system. By minimizing the residual, we can then iteratively converge to the true solution using a root-finding algorithm, such as the Newton-Raphson method. Our code uses an additional relaxation factor to improve convergence, as well as compactified coordinates to properly set boundary conditions, and adaptive mesh refinement near the boundaries. With all of this machinery, our code typically converges to the user prescribed tolerance in 1--3 iterations. 

After constructing this numerical infrastructure, we implement it in a few different scenarios. We begin by validating the algorithm through a simple toy problem whose analytic solution is known, and then through the calculation of the Schwarzschild solution in General Relativity. Since in both cases an analytic solution is known, we can easily compare it to our numerical solutions point-wise in the entire domain.  We find that our code converges to the correct (analytically known) solution in one and three iterations respectively to within the tolerance we specified. 

With the validation complete, we then construct stationary and spherically symmetric black hole solutions in scalar-Gauss-Bonnet (sGB) gravity, a well motivated modified theory that is a member of the quadratic gravity class~\cite{Yunes2013, 0264-9381-32-24-243001,Barack:2018yly}. Ours is of course not the first study of compact objects in sGB gravity. Mathematically, the theory evades the no-hair theorems of General Relativity, allowing black holes to have non-trivial scalar hair~\cite{CAMPBELL1992199,PhysRevD.47.5259,PhysRevD.54.5049,PhysRevD.83.104002, Sotiriou:2013qea,PhysRevD.90.124063}, while preventing neutron stars from having a monopole scalar charge~\cite{PhysRevD.93.024010}, thus making it extremely difficult to place constraints with binary pulsar observations.

Compact objects in scalar-Gauss-Bonnet gravity are typically studied under two forms of the action which depend on the coupling function between the massless scalar field and the Gauss-Bonnet invariant. When the scalar field is coupled through an exponential to the Gauss-Bonnet invariant this is commonly referred to as Einstein-dilaton-Gauss-Bonnet (EdGB) gravity. In the regime where the scalar field is small, one can approximate the exponential as a linear coupling to the Gauss-Bonnet invariant which is commonly referred to as the linear scalar-Gauss-Bonnet gravity. This is the terminology that will be used throughout this paper namely: `linear sGB' to refer to the linear coupling function and `EdGB' to refer to the exponential coupling function.

Stationary black holes have been found in linear sGB analytically using the small coupling limit approximation, both in spherical symmetry~\cite{PhysRevD.83.104002, Sotiriou:2013qea,PhysRevD.90.124063} and in axisymmetry using a slow-rotation approximation~\cite{CAMPBELL1992199, MIGNEMI1993299, PhysRevD.90.044066, PhysRevD.84.087501, PhysRevD.92.083014}. Reference~\cite{PhysRevD.90.124063} also studied numerical non-perturbative spherically symmetric black holes in linear sGB. In EdGB, numerical solutions are known in spherical symmetry~\cite{PhysRevD.54.5049, PhysRevD.55.739, PhysRevD.55.2110, BHinDilatonicEDGB, PhysRevD.79.084031} and in axisymmetry~\cite{PhysRevD.93.044047}, but typically these numerical solutions are obtained from a proprietary code that has not been tested for black hole problems of this kind. It is also worth mentioning that, dynamical evolution of black holes and binaries in sGB gravity has also received attention recently \cite{Benkel:2016kcq,Benkel:2016rlz,Witek:2018dmd,Ripley:2019hxt}.

Our numerical infrastructure is general enough to allow the tackling of any coupling function between the massless dilaton and the Gauss-Bonnet term in the action, and we here explore both the linear sGB and EdGB coupling functions. In the linear case, we find that the exact numerical solution agrees spectacularly well with the perturbative analytic solution that assumes weak-coupling, while differences arise in the exponential coupling case for large enough coupling. An example of this is shown in Fig.~\ref{fig:Masscomp}, where we present the relative fractional correction to the ADM mass (left) and the scalar monopole charge (right) as a function of the dimensionless sGB coupling parameter $\bar{\alpha} = \alpha/r_{\Hz}^2$. Although the differences between the EdGB and linear sGB solutions appears large for large $\bar{\alpha}$, the deviation is actually only appreciable near the horizon, as we will show later. 

With these exact numerical solutions at hand, we construct analytical fitted models to allow for the rapid computation of physical observables, such as the location of the innermost circular stable orbit (ISCO) and the light ring. The fitting coefficients are available online by request though we provide a few examples in Appendix~\ref{app:coeff}. Figure~\ref{fig:ISCOLRcomp} shows the fractional change in the location of the ISCO (left) and the light ring (right). We again find comparatively very good agreement between the analytic perturbative result and the exact numerical result of linear sGB for all $\bar{\alpha}$s considered, and some disagreement with the exact numerical result of EdGB for large $\bar{\alpha}$. Observe also that the observables computed with the exact numerical solution agree extremely well with those computed with our analytical fitted models. 
\begin{figure*}[htb]
\begin{center}
\resizebox{8cm}{!}{\include{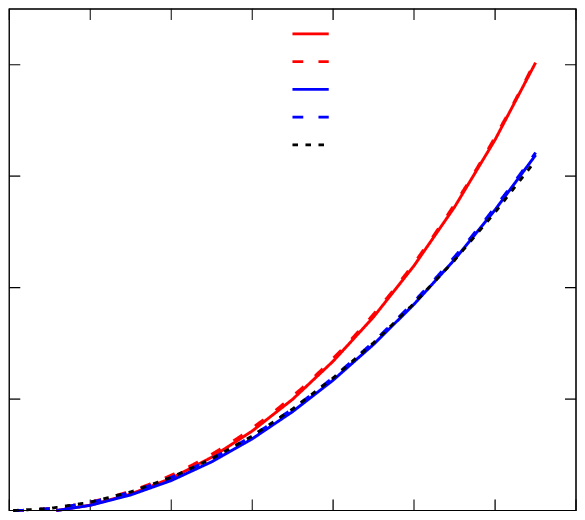}}
\resizebox{8cm}{!}{\include{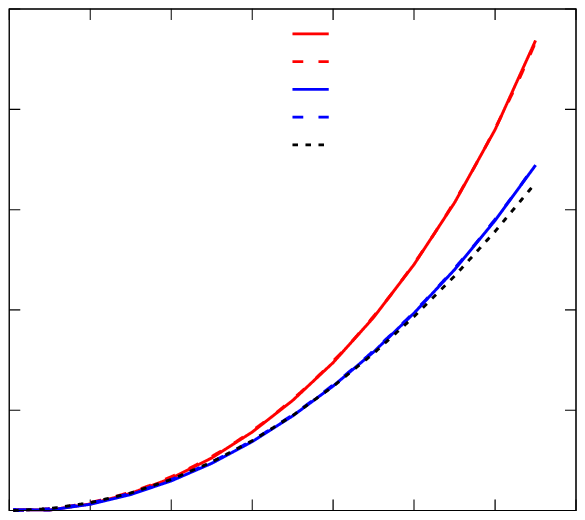}}
\caption{\label{fig:ISCOLRcomp} (Color Online) 
Relative fractional correction to the location of the innermost stable circular orbit (left) and the light ring (right). The red and blue lines correspond to these observables computed for EdGB or linear sGB respectively. The solid lines correspond to the observables computed with our exact numerical solution, the dashed lines use our analytical fitted model, and the black dotted lines use the analytical perturbative solution to linear order in the coupling.}
\end{center}
\end{figure*}

Our paper differs from previous work in that its goal is not to study a particular theory of gravity, but rather to develop a computational infrastructure that is (i) free and open to the public, (ii) computationally accurate and efficient, (iii) taylor-made for finding black hole solutions in modified theories of gravity with vector or scalar non-minimal couplings, and (iv) able to provide enough accurate data to compute analytical fitted models. The algorithm presented here applies to spherically symmetric scenarios, but extensions to axisymmetry are straightforward and under way. The algorithm developed here can thus be used to calculate certain astrophysical observables, such as the electromagnetic emission of accretion disks around black holes~\cite{Abramowicz2013}, the shadows of black holes~\cite{0264-9381-35-23-235002}, or the quasinormal modes of black hole mergers~\cite{Blazquez-Salcedo:2016enn} on the fly and with high precision, allowing for realistic data analysis investigations with Bayesian methods.

The remainder of this paper is organized as follows. 
Section~\ref{sec:NM} outlines the numerical algorithm as applied to a general system of equations. 
Section~\ref{sec:GR} validates the algorithm through a simple toy problem and a Schwarzschild black hole. 
Section~\ref{sec:EdGB} applies the algorithm to sGB gravity and derives the results described above. 
Section~\ref{sec:props} constructs a fitted analytical model from the numerical solutions and compares physical observables determined by the numerical solutions and the fits. 
Finally, Section~\ref{sec:concs} summarizes our results and points to future directions. 
For the remainder of this paper we use the following conventions: Greek letters denote spacetime indices; the metric has the spacetime signature $\lb -,+,+,+ \rb$; we use geometric units where $G = 1 = c$.

\section{Numerical Methods}
\label{sec:NM}

Our numerical algorithm extends the method presented in~\cite{SCHONAUER1989279} to build a partial differential equation solver that finds exact black hole solutions in an arbitrary modified theory of gravity. The algorithm is split into three main parts: the relaxed Newton-Raphson method, the discretization method, and the discretization error estimation. We begin by outlining the Newton-Raphson method in the continuum limit in Sec.~\ref{ssec:NewtonsMethod}. The discretization method takes our generic system of nonlinear differential equations and recasts them for evaluation on a discrete domain through a Newton polynomial scheme outlined in Sec.~\ref{ssec:Newtpoly}. This method naturally introduces errors that must be estimated and controlled, as outlined in Sec.~\ref{ssec:mindiscerror}. Combining these results, we describe the application of the discrete Newton-Raphson method with a relaxation factor in Sec.~\ref{ssec:minres}, which reduces the computational problem to solving a system of coupled linear equations through two iterative Krylov subspace methods, detailed in Sec.~\ref{ssec:LS}. 

Given the description above, clearly this section contains a considerable amount of descriptive detail about the numerical methods that constitute the computational infrastructure we have deployed to find numerical black hole solutions in modified gravity. Our hope is that these details will serve as a sort of tutorial for theoretical physics experts that are less familiar with the numerical methods we use here. Readers already familiar with these numerical methods may wish to skip this section altogether and jump directly to the validation of the algorithm and its application to black holes in sGB in Sec.~\ref{sec:GR} and beyond. 

\subsection{Iterative Relaxation Method in the Continuum Limit}
\label{ssec:NewtonsMethod}

We start with a system of $M$, second-order, nonlinear elliptic partial differential equations. In the simplest 1-D (spherically symmetric and stationary) case, these equations are ordinary differential equations of some independent variable $x$, such that in operator form
\be
\label{eq:Dust}
\mcl{D} \, {\vec{u}}^{\,\ast}  = 0\,,\\
\ee
where $\mcl{D}$ is an $M \times M$ matrix of nonlinear differential operators and ${\vec{u}}^{\,\ast}$ is the \emph{solution vector}\footnote{For the remainder of this paper, the word \emph{vector} stands for a standard Euclidean vector in flat space.} with $M$ elements. The elements of $\vec{u}$ are the fields in the problem, which we will sometimes denote with capital latin indices $\vec{u} = (u_{0}, u_{1}, \ldots, u_{A}, \ldots, u_{M-1})$. We see then that the special solution vector ${\vec{u}}^{\,\ast}$ is annihilated by the differential operator $\mcl{D}$.

In general, however, one does not know the solution to the differential system, so a generic vector ${\vec{U}}$ will not be annihilated by $\mcl{D}$. Rather, for a generic vector ${\vec{U}}$ that is not a solution to the differential system one has
\be
\label{eq:Diffeq}
\mcl{D} \, {\vec{U}}  = \vec{b} (\vec{U}), \\
\ee
where the \emph{residual vector} ${\vec{b}}$ has $M$ elements and is a function of $\vec{U}$. The generic vector $\vec{U}$ is a functional that spans the vector function space that contains the solution vector $\vec{u}^{\, \ast}$ and any other vector that does not satisfy the differential system $\vec{u}$\footnote{In particle physics, the set of functions that satisfy a differential system of equations is sometimes referred to as ``on shell," while those that do not are referred to as ``off shell." Thus $\vec{U}$ spans the space that contains both on and off shell functions.}. Clearly then, once one finds the correct vector ${\vec{U}}$, namely ${\vec{U}} = {\vec{u}}^{\,\ast}$, then $\vec{b} (\vec{u}^{\, \ast}) = 0$. 

The vector ${\vec{U}}$ is also subject to boundary conditions. We can express these in the form
\be
\label{eq:BCeq}
\mcl{B} \, {\vec{U}} |_{\pd V} = {\vec{B}},
\ee
where $\mcl{B}$ is another $M \times M$ matrix typically composed of a combination of real numbers and first-order differential operators, while $\pd V$ is the boundary of the spatial domain. The \emph{boundary vector} ${\vec{B}}$ also has $M$ elements and is a set of constants defined on the boundary $\pd V$. Defining the boundary conditions in this way allows the imposition of Neumann boundary conditions for a subset of the components of $\vec{U}$ and Dirichlet boundary conditions for other components.

Let us assume that we have a vector $\vec{u}$ that we know is close the the actual solution vector $\vec{u}^{\, \ast}$. If so, the difference between it and the actual solution is some other small vector $\Delta \vec{u}$. Let us refer to the latter as the \emph{correction vector}, which is mathematically defined via
\be
\label{eq:ucorr}
\Delta \vec{u} := \vec{u}^{\, \ast} - \vec{u}.
\ee
We begin by linearizing $\vec{b}(\vec{U})$ around $\vec{U}=\vec{u}$ through a first-order Taylor expansion, analogous to the familiar Taylor expansion of $f(x)$ about $x = a$. Doing so, we find
\be
\label{eq:btaylor}
\vec{b} (\vec{U}) \approx \vec{b} (\vec{u}) + \left.\frac{\pd \vec{b}}{\pd \vec{U}} \right|_{\vec{u}} \lb \vec{U} - \vec{u} \rb.
\ee
We then evaluate this expression at the solution vector $\vec{U} = \vec{u}^{\, \ast}$ and obtain
\be
\vec{b} (\vec{u}^{\, \ast}) \approx \vec{b} (\vec{u}) + \left.\frac{\pd \vec{b}}{\pd \vec{U}} \right|_{\vec{u}} \lb \vec{u}^{\, \ast} - \vec{u} \rb.
\ee
Using that $\vec{b} (\vec{u}^{\, \ast}) = 0$ and substituting the correction vector defined in Eq.~\eqref{eq:ucorr}, this can be simplified to
\be
\label{eq:Newt}
{\mcl{J}} \Delta \vec{u} \approx - \vec{b} (\vec{u}),
\ee
where we have defined the $M \times M$ Jacobian matrix
\be
{\mcl{J}} := \left.\frac{\pd \vec{b}}{\pd \vec{U}}\right|_{\vec{u}}\,.
\ee

We now wish to solve for the correction vector $\Delta \vec{u}$, which implies we must either invert the Jacobian or solve the linear system in Eq.~\eqref{eq:Newt}. In practice, matrix inversion is typically more computationally expensive than linear system solving, so the latter is the method we use here, which we will explain in more detail in Sec.~\ref{ssec:LS}. If one knew the correction vector $\Delta \vec{u}$ exactly, one could then find the solution vector from Eq.~\eqref{eq:ucorr}. But in reality the linearizaton in Eq.~\eqref{eq:btaylor} implies the correction vector we find by solving Eq.~\eqref{eq:Newt} is only an approximation to the true correction vector. This implies that to find the true correction vector we must apply this procedure \emph{iteratively}.

Let us then describe the first couple of iterations of this procedure. We start with an initial guess $\vec{u}^{(0)}$ that we know is close to the true solution, where the superscript in parenthesis is the iteration number. In this paper, the initial guess can be chosen to be either the GR solution, or an approximate solution for all fields in the modified theory. With this initial guess, we then find the initial residual vector
\be
\vec{b}^{(0)} = \mcl{D} \vec{u}^{(0)}.
\ee
Since the initial guess is not a solution to the differential system, the residual vector does not vanish, and we must thus correct the initial guess to find the first-iterated solution
\be
\vec{u}^{(1)} = \vec{u}^{(0)} + \Delta \vec{u}^{(0)}\,.
\ee
This requires the calculation of the zeroth-iterated correction vector $ \Delta \vec{u}^{(0)}$, which we find by solving the linear system
\be
\label{eq:Newt-0th}
{\mcl{J}}^{(0)} \Delta \vec{u}^{(0)} = - \vec{b} (\vec{u}^{(0)})\,,
\ee
where the Jacobian is evaluated on the initial guess $\vec{u}^{(0)}$. This procedure then yields $\vec{u}^{(1)}$, and now it can be repeated until the n\textsuperscript{th}-iteration to the solution $\vec{u}^{(n)}$ is sufficiently close to $\vec{u}^{\, \ast}$, i.e.~until the residual vector is below some specified tolerance $\vec{b} (\vec{u}^{(n)}) < \mrm{tol}$.

\subsection{Discrete Representation through Newton's Polynomials}
\label{ssec:Newtpoly}

In order to numerically solve the differential system described above, one first needs to discretize it on a finite numerical grid. We here use Newton's (centrally divided difference) interpolation polynomial method, which we describe next. 

Newton's interpolation polynomial provides a continuous local representation of a discrete function given by a set of data points. This procedure is a discrete analog of using a Taylor series to represent an approximation to a continuous function $f(x)$ as a local polynomial about some point $x = a$, namely
\bal
\label{eq:taylorseries}
\bsp
f(x) ={}& \sum_{n=0}^{\infty} \frac{f^{(n)}(a)}{n!} \lb x - a \rb^n, \\
={}& f(a) + f^{\prime}(a) \lb x - a \rb + \frac{1}{2} f^{\prime \prime} (a) \lb x - a \rb^2 + \ldots,
\esp
\end{align}
where the primes denote derivatives with respect to the independent variable $x$. 

Now imagine that instead of a continuous function $f(x)$, we have a discrete function ${u}_d (x)$ known only on a discrete collection of $N$ data points $[(x_0, u_0), (x_1, u_1), \ldots, (x_i, u_i),\ldots, (x_{N-1}, u_{N-1})]$. Notice that the lower-case Latin subscript here does not denote the components of the $\vec{u}$ of the previous subsection, but rather the element $x_{i}$ at which we evaluate the discrete function ${u}_{d}$. For notational convenience, we identify $u_i$ with the discrete function $u_d(x)$ evaluated at each point $x_i$, namely
\be
\label{eq:u_idef}
u_i = u_d(x_i).
\ee

How do we now approximate the discrete function ${u}_{d}$ in a neighborhood of some point in the spatial domain? We are tempted to use a Taylor expansion again, but because our function is discrete, we cannot take analytical derivatives as we did before, and instead we must use a finite difference approximation. Let us then define a neighborhood around some point $x_{I}$ as the region in the discrete spatial domain around which we wish to approximate our function. Clearly then, the value of the function at the point $x_{I}$ is simply $u_I = u_d(x_{I})$. Let us further temporarily assume that the data points are equidistant, such that $\Delta x = x_{I+1} - x_{I}$. 

We are now almost ready to define our finite difference approximation, but first we must choose which discrete points in the neighborhood of $x_{I}$ we will use to approximate the function. For example, if we choose to use the points $(x_{I}, x_{I+1}, x_{I+2})$, then the Newton interpolation polynomial of our discrete function is
\bal
\label{eq:Newtsimpforward}
\bsp
u_{d}(x) &= u_{I} + \frac{u_{I+1} - u_{I}}{\Delta x} \lb x - x_{I} \rb \\
+{}& \frac{u_{I+2} - 2 u_{I+1} + u_{I}}{2 \Delta x^2} \lb x - x_{I} \rb \lb x - x_{I+1} \rb.
\esp
\end{align}
The coefficients of the second and third terms in Eq.~\eqref{eq:Newtsimpforward} are simply the first- and second-order forward finite difference approximation to the first and second derivative of the function in the $\Delta x \rarr 0$ limit, namely
\bal
\label{eq:Newtsimpforwardfirst}
\bsp
u_d^{\prime}(x_{I}) \equiv{}& \lim_{\Delta x \rarr 0} \frac{u_d(x_{I} + \Delta x) - u_d(x_{I})}{\Delta x}, \\
={}& \frac{u_d(x_{I} + \Delta x) - u_d(x_{I})}{\Delta x} + \mcl{O}(\Delta x).
\esp
\end{align}
and
\bal
\bsp
{}& u_d^{\prime \prime}(x_{I}) \equiv \\
& \lim_{\Delta x \rarr 0} \frac{u_{d}(x_{I}+2 \Delta x) - 2 u_d(x_{I}+ \Delta x) + u_d (x_{I})}{\Delta x^2}, \\
& = \frac{u_{d}(x_{I}+2 \Delta x) - 2 u_d(x_{I}+ \Delta x) + u_d (x_{I})}{\Delta x^2} \\
& + \mcl{O}(\Delta x),
\esp
\end{align}
respectively. Moreover, using that 
\be
\label{eq:basiscorrection}
(x - x_{I}) (x - x_{I+1}) = \lb x - x_{I} \rb^2 - \Delta x \lb x - x_{I} \rb,
\ee
we notice that the $(x-x_{I})^2$ term is the same polynomial that appears in the Taylor expansion, and that the additional $\Delta x$ term cancels with one of the $\Delta x^2$ in the denominator of Eq.~\eqref{eq:Newtsimpforward}. We can then combine this with the first-order term to obtain
\bal
\bsp
{}& u_d^{\prime}(x_{I}) = \\
& \frac{-u_d (x_{I} + 2 \Delta x) + 2 u_d (x_{I} + \Delta x) - 3 u_d (x_{I})}{2 \Delta x} \\
&+ \mcl{O}(\Delta x^2).
\esp
\end{align}

Let us now compare the above result to the first-order finite difference derivative in Eq.~\eqref{eq:Newtsimpforwardfirst}. First, the derivative now includes three points instead of two. Second, the accuracy has been increased to $\mcl{O}(\Delta x^2)$. This suggests that as we add increasingly higher-order polynomials to our discrete representation, they will automatically include corrections to each of the previous lower-order polynomials. Thus, the Newton interpolation polynomial is simply a discrete analog to the Taylor series, and they are formally equivalent in the continuous limit.

The full construction of the Newton interpolation polynomial requires certain Newton basis polynomials and certain divided differences coefficients, which are a generalization of the finite difference coefficients of the above simple example. The choice of points to include leads to the distinction between a forward and a backward divided difference, and this influences the directional finite difference that reduces to the Taylor derivative in the continuum limit. As a simple example, consider again the approximation of a discrete function $u_{d}$ in a neighbourhood around a point $x_{I}$ but this time using the points immediately surrounding $x_{I}$: $(x_{I}, x_{I+1}, x_{I-1})$. Equation~\eqref{eq:Newtsimpforward} then becomes
\bal
\label{eq:Newtsimpcentral}
\bsp
u_{d}(x) &= u_{I} + \frac{u_{I+1} - u_{I}}{\Delta x} \lb x - x_{I} \rb \\
+{}& \frac{u_{I+1} - 2 u_{I} + u_{I-1}}{2 \Delta x^2} \lb x - x_{I} \rb \lb x - x_{I+1} \rb,
\esp
\end{align}
which results in the first and second-order \emph{central} finite difference equations in the continuum limit. Similarly if we use the points before $x_{I}$ only, namely $(x_{I}, x_{I-1}, x_{I-2})$, we then obtain the \emph{backwards} finite difference equations in the continuum limit. The accuracy of the central finite difference equations is improved by a factor of $\mcl{O}(\Delta x)$ over the forward or backwards finite differences.

Let us generalize the systematic application of these choices through the introduction of a \emph{stencil}, namely a Euclidean vector $s_{j}$ whose elements are the labels of the points included in the evaluation of our Newton polynomial. For example, for a forward, central, and backwards Newton polynomial representation of a discrete function about a point $x_{I}$, the stencil $s_{j} = (I, I+1, I+2), (I, I+1, I-1),$ and $(I, I-1, I-2)$ respectively. 

With this at hand we can write the general Newton interpolation polynomial of a discrete function $u_{d}$ about any point $x$ in the spatial domain as
\be
\label{eq:Newt-pol}
u_{d}(x) = \sum \limits_{j=0}^{r} \sum \limits_{i=0}^{j} u_{s_i} \; w_{s_i,s_j} \; P_{s_j} (x),
\ee
where recall that $u_{s_i}$ is given by Eq.~\eqref{eq:u_idef} using stencil notation, and where
\be
w_{s_{i}, s_{j}} = \frac{1}{\prod \limits_{\substack{p = 0 \\ p \neq i}}^{j} \lb x_{s_{i}}-x_{s_{p}} \rb},
\ee
and $P_{s_j} (x)$ are the Newton basis polynomials, defined by
\be
\label{eq:basis}
P_{s_j} (x) = \prod \limits_{q=0}^{j-1} \lb x-x_{s_q} \rb.
\ee
If the grid is uniform and $\Delta x = x_{I+1} - x_{I}$, one can check that this equation reduces to Eqs.~\eqref{eq:Newtsimpforward} or~\eqref{eq:Newtsimpcentral} with the respective stencil.

The index $r$ in Eq.~\eqref{eq:Newt-pol} is the order of the Newton polynomial and it indicates how many grid points are used and the maximum finite difference derivative order. In practice, as the order $r$ increases, we successively add points to either side of $x_{I}$ to keep the coefficients as central as possible, so our stencil has the form $s_j = ({I}, {I+1}, {I-1}, {I+2}, {I-2}, \ldots)$ and $s_j$ will have $r+1$ elements. For even $r$ the stencil will be purely central, whereas for odd $r$ the stencil will be slightly forward. On the boundary of our domain, we must add points in a one-sided manner (either forward or backwards), and we must add an extra point beyond the number of points we keep away from the boundaries, to keep the accuracy of the represetation comparable to the central differences. Note that the specific sequence of the elements of $s_j$ does not change the resulting polynomial. 

\begin{figure}[htb]
\begin{center}
\resizebox{8cm}{!}{\include{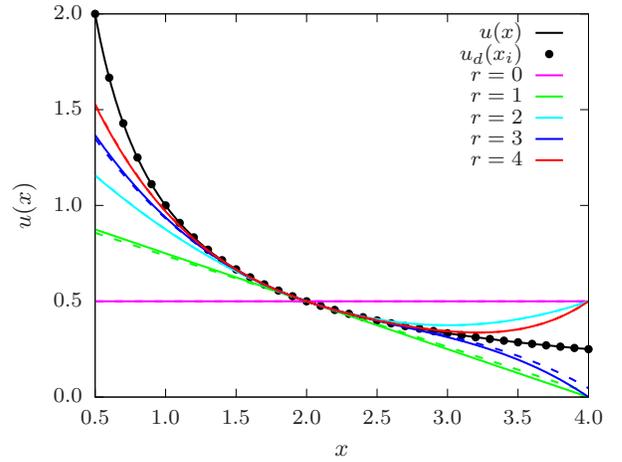}}
\caption{\label{fig:NewtTaylorcomp} (Color Online) 
Taylor series and Newton interpolation polynomial representation of the toy function $u(x) = 1/x$ around the point $x_{I} = 2$ on a uniform discretized grid order by order. The black solid line is the analytical function and the black dots are the discretized data points with $\Delta x = 0.1$. Solid lines indicate the Taylor series and dashed lines indicate the Newton polynomial, while the different colors indicate the polynomial order $r$. Observe the improved approximation for the even $r$ orders compared to the odd-orders due to the slightly one-sided nature of the stencil for odd-orders.}
\end{center}
\end{figure}

As an example of an application of this, let us calculate and compare the Newton interpolation polynomial representation and the comparable Taylor series expansion of a toy function 
\be
u(x) = 1/x,
\ee
on a uniform discretized grid where $\Delta x = 0.1$ around the point $x_{I} = 2$. With a maximum order $r = 4$, the centralized stencil is $s_j = (I, I+1, I-1, I+2, I-2)$. Figure~\ref{fig:NewtTaylorcomp} shows the Taylor series expansion and the Newton polynomial representation of our toy function order by order. The first three terms are shown in Eqs.~\eqref{eq:taylorseries} and~\eqref{eq:Newtsimpcentral}, from which we see the $r = 0$ order approximation in both cases is simply a constant $u_{d}(2) = 1/2$. The $r = 1$ order is a line whose slope is calculated by either the derivative of the function evaluated at $x_{I} = 2$, or the first-order finite difference evaluated at $x_{I}=2$. Observe that the agreement between the Newton interpolation polynomial representation of the discrete function and the Taylor series expansion of the continuous function is very close but the agreement is better for even-orders of $r$ than for odd-orders. This is a result of the even-orders of $r$ being purely central for example for $r = 2, s_{j} = (I, I+1, I-1)$ while the odd-orders are slightly forward for example $r = 3, s_{j} = (I, I+1, I-1, I+2)$ which diminishes the accuracy. It is for this reason that we will restrict our choice in $r$ to be even.
%

\subsection{Derivatives of Newton Polynomials and Discretization Error}
\label{ssec:dis-error}

One of the main advantages of using a Newton interpolation polynomial representation of a discrete function is that with it one can take \emph{analytic} derivatives of the discrete function. For example, the derivatives $\pd_x u_{d} := {d u_{d}}/{d x}$ can obtained by taking an analytic derivatives of the basis functions $P_j$, which are the only $x$ dependent terms in Eq.~\eqref{eq:Newt-pol}. The two relevant derivatives for second order equations are
\be
\pd_{x} P_j (x) = \sum \limits_{k=0}^{j-1} \prod \limits_{\substack{i=0 \\ i \neq k}}^{j-1} \lb x-x_i \rb,
\ee
and
\be
\pd_{xx} P_j (x) = \sum \limits_{l=0}^{j-1} \sum \limits_{\substack{k=0 \\ k \neq l}}^{j-1} \prod \limits_{\substack{i=0 \\ i \neq k \\ i \neq l}}^{j-1} \lb x-x_i \rb,
\ee
as one can verify by direct differentiation. 

This property of Newton interpolation polynomial representations allows us to replace the action of any differential operator on our fields at any grid point with a finite difference coefficient coupled to the $r$ neighboring grid points, but this introduces some error. To quantify this error, let us introduce boldface vector notation to denote a \emph{discretized vector}, i.e.~the collection of data points of the Newton polynomial representation of a field evaluated on every point of the discretized grid. For example, for a given field $\bm{u}_{d} = [u_{d}(x_{0}), u_{d}(x_{1}), \ldots, u_{d}(x_{N})]$. By construction, of course, the discretized field itself has no discretization error on any point of the discretized domain, but this is not the case for derivatives of the field. Let us then introduce the discretization error vector $\bm{e}_d$ and the discretization operator $\rarr$, such that
\bal
\label{eq:fulldisc}
\bsp
u &\rarr \bm{u}_d, \\
\pd_{x} u &\rarr \pd_{x} \bm{u}_d + \pd_{x} \bm{e}_d, \\
\pd_{xx} u &\rarr \pd_{xx} \bm{u}_d + \pd_{xx} \bm{e}_d, \\
\esp
\end{align}
because by construction $\bm{e}_d (x_{i}) = 0$. These discretized errors will be incorporated into the Newton's method of Sec.~\ref{ssec:mindiscerror} to maintain numerical accuracy. 

Before moving on to the next subsection, let us here address some potential confusion due to the different choices of notation for a vector that we have employed. The boldface vector notation of the previous paragraph denotes a \emph{discretized vector}, so its components are the values of the Newton polynomial representation of a function on each grid point of our spatial domain. The arrow vector notation of the previous section denotes a \emph{generic field vector}, so its components are the different fields in the differential system. When we return to our general system of $M$ differential equations, we must generate a Newton polynomial for each component of $\vec{u}$, and then, we must discretize each of these representations on $N$ grid points. The quantity $\vec{\bm{u}}_d$ then becomes a 2-D matrix (see Eq.~\eqref{eq:notationnightmare} below) with each element denoted as $u_{A,i}$. 

As we will see later, we will have to \emph{fold} every discretized vector in our system of equations so that we can take proper element by element partial derivatives. To do this, we introduce a new index $k = i N + A$ and fold the $\vec{\bm{u}}_d$ matrix into a single vector where $k \in [0, (M-1)(N-1)]$, namely
\bw
\bal
\label{eq:notationnightmare}
\bsp
\vec{u} = 
\begin{bmatrix}
u_{0} \\
\vdots \\
u_{A} \\
\vdots \\
u_{M-1} \\
\end{bmatrix}
\rarr
\vec{\bm{u}}_d ={}&
\begin{bmatrix}
\bm{u}_{0} \\
\vdots \\
\bm{u}_{A} \\
\vdots \\
\bm{u}_{M-1} \\
\end{bmatrix}
= 
\begin{bmatrix}
u_{0,0} & \ldots & u_{0,i} & \ldots & u_{0,N-1}\\
\vdots \\
u_{A,0} & \ldots & u_{A,i} & \ldots & u_{A,N-1}\\
\vdots \\
u_{M-1,0} & \ldots & u_{M-1,i} & \ldots & u_{M-1,N-1}\\
\end{bmatrix}, \\
={}& \lsb u_{0,0}, \ldots, u_{A,0}, \ldots, u_{M-1,0}, \ldots, u_{0,i}, \ldots, u_{A,i}, \ldots, u_{M-1,i}, \ldots, u_{0,N-1}, \ldots \rsb, \\
={}& \lcb u_{k} \rcb, \qquad k \in [0, (N-1)(M-1)].
\esp
\end{align}
\ew
This folding replaces our system of $M$ continuous differential equations with a linear system of $M \times N$ equations. 

\subsection{Minimization of the Discretization Error}
\label{ssec:mindiscerror}

Let us consider how the discretization error in the derivatives of the Newton polynomial representation of our discrete fields propagates into our system of $M$ differential equations. For a generic vector $\vec{U}$, our discretized differential system becomes
\be
\label{eq:disc-residual-eq}
\mcl{D} \vec{U} = \vec{b}(\vec{U}) \rarr \bm{\mcl{D}} \lb \vec{\bm{U}}_d + \vec{\bm{e}}_d \rb = \vec{\bm{b}}_d (\vec{\bm{U}}_d) + \tilde{\bm{\mcl{D}}} \vec{\bm{e}}_d, \\
\ee
where $\tilde{\bm{\mcl{D}}}$ is the linearized differential operator resulting from the expansion of $\mcl{D}$. As before, let us choose our vector $\vec{u}$ to be close to the solution vector $\vec{u}^{\, \ast}$ up to a small correction vector $\Delta\vec{u}$. Discretizing these relations we then have that
\be
\vec{u} = \vec{u}^{\, \ast} - \Delta \vec{u} \rarr \vec{\bm{u}}_{d} = \vec{\bm{u}}^{\, \ast}_{d} - \Delta \vec{\bm{u}}_{d} - \Delta \vec{\bm{u}}_e,
\ee
where we have included a new correction vector to attempt to also minimize the discretization error. As in Sec.~\ref{ssec:NewtonsMethod}, we then Taylor expand Eq.~\eqref{eq:disc-residual-eq} to find
\be
\label{eq:Newterror}
{\bm{\mcl{J}}} \lb \Delta \vec{\bm{u}}_d + \Delta \vec{\bm{u}}_e \rb = - \lb \vec{\bm{b}}_d + \vec{\bm{D}}_e \rb,
\ee
where the discretized Jacobian is ${\bm{\mcl{J}}} = {\pd \vec{\bm{b}}_d}/{\pd \vec{\bm{u}}_d}$ evaluated at $\vec{\bm{u}}_{d}$, and we have defined the \emph{discretization error vector} $\vec{\bm{D}}_e \equiv \tilde{\bm{\mcl{D}}} \vec{\bm{e}}_d$. The above equation can be satisfied by separately and simultaneously requiring that 
\be
\label{eq:Newterror-1}
{\bm{\mcl{J}}} \Delta \vec{\bm{u}}_d = - \vec{\bm{b}}_d,
\ee
for the discretized correction vector $\Delta \vec{\bm{u}}_d$  and
\be
\label{eq:Newterror-2}
{\bm{\mcl{J}}} \Delta \vec{\bm{u}}_e  = - \vec{\bm{D}}_e,
\ee
We recognize Eq.~\eqref{eq:Newterror-1} as the discretized version of Eq.~\eqref{eq:Newt}. By balancing both of these equations we can monitor the numerical accuracy of our discretization error.

The equation for the discretization error vector, Eq.~\eqref{eq:Newterror-2}, describes the change to our solution due to the discretized error vector $\vec{\bm{D}_e}$. To control the discretized error, we wish to require that the relative correction is below a specified tolerance. To solve Eq.~\eqref{eq:Newterror-2}, we first must calculate $\vec{\bm{D}}_e$ which was defined as the discretized differential operator acting on $\vec{\bm{e}}_d$:
\be
\label{eq:Ddef}
\vec{\bm{D}}_e \equiv \tilde{\bm{\mcl{D}}} \vec{\bm{e}}_d.
\ee
In practice, we will generally not have our system of nonlinear equations in operator form. For example, when calculating the Einstein field equations we will have the set of equations in the form of the residual vector $\vec{\bm{b}}(\vec{\bm{u}})$ as a functional of the fields themselves. If we wish to extract the linearized differential operator of Eq.~\eqref{eq:Ddef} we must calculate it from the residual vector. 

As a simple example, consider the following single continuous nonlinear differential equation in operator notation:
\be
\mcl{D} u = \lsb \pd_{xx} u + \lb \pd_{x} u \rb^2 \rsb.
\ee
If we wish to calculate $u \rarr \vec{u}_d + \vec{e}_d$ we obtain,
\bal
\bsp
\bm{\mcl{D}} \lb \vec{u}_d + \vec{e}_d \rb ={}& \pd_{xx} \vec{u}_d + \pd_{xx} \vec{e}_d + \lb \pd_{x} \vec{u}_d \rb^2 \\
+{}& 2 \pd_{x} \vec{u}_d \pd_{x} \vec{e}_d + \lb \pd_{x} \vec{e}_d \rb^2.
\esp
\end{align}
From this, we identify the linearized operator as
\be
\tilde{\bm{\mcl{D}}} \vec{e}_d = \pd_{xx} \vec{e}_d + 2 \pd_{x} \vec{u}_d \pd_{x} \vec{e}_d,
\ee
and the general discretization error vector as
\be
\label{eq:De}
\vec{\bm{D}}_e = \frac{\pd \vec{\bm{b}}_d}{\pd \lb \pd_{x}\vec{\bm{u}}_d \rb} \pd_{x} \vec{\bm{e}}_d + \frac{\pd \vec{\bm{b}}_d}{\pd \lb \pd_{xx} \vec{\bm{u}}_d \rb} \pd_{xx} \vec{\bm{e}}_d,
\ee
where as before $\vec{\bm{e}}_d = 0$ due to the Newton polynomial construction. 

The derivative error can then be computed from Eq.~\eqref{eq:fulldisc}, namely via
\bal
\bsp
\pd_{x} \vec{\bm{e}}_d ={}& \pd_{x} \vec{u} - \pd_{x} \vec{\bm{u}}_d, \\
\pd_{xx} \vec{\bm{e}}_d ={}& \pd_{xx} \vec{u} - \pd_{xx} \vec{\bm{u}}_d, \\
\esp
\end{align}
but in general we do not know the analytical derivatives of any of the fields $u$. Following~\cite{SCHONAUER1989279}, we estimate the discrete derivative error from the difference between the discretized derivative at order $r$ and that at order $r+2$. This is valid as long as the error decreases for increasing order, and if so, it provides a reasonable estimate as demonstrated in~\cite{SCHONAUER1981327}. With this approximation, our discretized derivative error is
\bal
\bsp
\label{eq:ped}
\pd_{x} \vec{\bm{e}}_d ={}& \pd_{x} \vec{\bm{u}}_d^{(r+2)} - \pd_{x} \vec{\bm{u}}_d^{(r)}, \\
\pd_{xx} \vec{\bm{e}}_d ={}& \pd_{xx} \vec{\bm{u}}_d^{(r+2)} - \pd_{xx} \vec{\bm{u}}_d^{(r)}. \\
\esp
\end{align}

Let us then describe the procedure we will follow to solve Eq.~\eqref{eq:Newterror}. First, before finding the field correction to the zeroth-iteration,  we ensure the discretization error is under a specified tolerance by using Eq.~\eqref{eq:Newterror-2} to determine the discretized grid size. This is achieved through a separate Newton-Raphson subroutine that we describe below. Once this is achieved, we then solve Eq.~\eqref{eq:Newterror-1} to find the zeroth-iteration correction vector, which allows us to update the solution. This full procedure is then iterated, at each step ensuring that the discretization error is under control, until the residual in Eq.~\eqref{eq:Newterror-1} is below a prescribed tolerance. We will detail this process in Sec.~\ref{ssec:minres}.

Let us now describe the Newton-Raphson subroutine that we use to minimize the discretization error. First, we calculate $\vec{\bm{D}}_e$ from Eq.~\eqref{eq:De} using Eq.~\eqref{eq:ped} for the discretization derivative error.   We then find the correction to the solution $\Delta \vec{\bm{u}}_e$ by solving
\be
{\bm{\mcl{J}}} \Delta \vec{\bm{u}}_e = - \vec{\bm{D}}_e,
\ee
using one of the iterative methods in Sec.~\ref{ssec:LS}. We stop the subroutine when the following condition is satisfied
\be
\label{eq:adstepcond}
\frac{\left\Vert \Delta \vec{\bm{u}}_e \right\Vert}{\left\Vert \vec{\bm{u}}_{d} \right\Vert} \leq \mrm{tol},
\ee
where $\left\Vert \cdot \right\Vert$ is the maximum norm (also called infinity norm or supremum norm) among all fields and grid points. This condition forces the relative correction to the solution from the discretization error to be below some specified tolerance $\rm{tol}$. In turn, this ensures that our final solution is within the discretized error tolerance.

If the above condition is not met for the chosen grid discretization, we adaptively adjust the grid step size until the relative discretization error correction is below the specified tolerance. To adjust the step size at each grid point $i$ defined as $\Delta x_{i}^{\rm{old}}$, we rescale the latter by the ratio of Eq.~\eqref{eq:adstepcond} with an additional ``safety'' factor of $1/3$, and relax the ratio by the Newton polynomial order $r$, such that
\be
\label{eq:dxnew}
\Delta x_{i}^{\new} = \lb \frac{\frac{1}{3} \left\Vert \vec{\bm{u}}_{d} \right\Vert \mrm{tol}}{\left| \Delta \vec{\bm{u}}_d \right|_{i}} \rb^{\frac{1}{r}} \Delta x_{i}^{\old},
\ee
where $\left| \cdot \right|_{i}$ is the maximum norm among all fields evaluated at grid index $i$. 

Once the new step sizes are calculated, the entire grid must be adjusted and the condition in Eq.~\eqref{eq:adstepcond} must be rechecked. In the grid adjustment, we must ensure that the location of a new grid point remains within half an old step size of the old grid point. Then, each function is interpolated using the Newton polynomial representation calculated from the old grid point locations. From the new interpolated solution, $\vec{\bm{D}}_e$ and $\Delta \vec{\bm{u}}_e$ are recalculated and the condition of Eq.~\eqref{eq:adstepcond} is rechecked. If the condition is not satisfied, the process using Eq.~\eqref{eq:dxnew} is repeated until the step sizes on the grid become sufficiently small that the condition is satisfied. Once the discretization error is sufficiently under control we are ready to finally apply the Newton-Raphson iterations to the solution vector.

\subsection{Minimization of the Residual Vector}
\label{ssec:minres}

As long as we can ensure our discretization error is under control, we can safely ignore their contributions to Eq.~\eqref{eq:Newterror} and we are left with the equation
\be
\label{eq:fullnewt}
{\bm{\mcl{J}}} \Delta \vec{\bm{u}}_d = - \vec{\bm{b}}_d\,,
\ee
which we can now solve following the same steps outlined in Sec.~\ref{ssec:NewtonsMethod}. 

The full algorithm is then as follows. Using the system at iteration $n$, $\vec{\bm{u}}_d^{(n)}$, we first find the discretized residual vector
\be
\vec{\bm{b}}_d^{(n)} = \bm{\mcl{D}} \vec{\bm{u}}_d^{(n)},
\ee
and calculate the discretization error $\vec{\bm{D}}_d^{(n)}$. We then check Eq.~\eqref{eq:adstepcond} to determine if we need to adjust the step size according to Eq.~\eqref{eq:dxnew}. Once the step size is sufficient, we solve Eq.~\eqref{eq:fullnewt} for the correction vector $\Delta \vec{\bm{u}}_d^{(n)}$ and the generic vector is updated with a relaxation factor $\omega$ (initially $\omega = 1$) added
\be
\vec{\bm{u}}_d^{(n+1)} = \vec{\bm{u}}_d^{(n)} + \omega \Delta \vec{\bm{u}}_d^{(n)}.
\ee
Before the new solution is accepted, the residual of the system is re-calculated to ensure convergence, i.e.~that the new residual is smaller than the old residual:
\be
\label{eq:convcond}
\left\Vert \vec{\bm{b}}_{d}^{(n+1)} \right\Vert < \left\Vert \vec{\bm{b}}_{d}^{(n)} \right\Vert.
\ee
If the convergence condition of Eq.~\eqref{eq:convcond} does not hold then the relaxation factor is reduced by half,
\be
\omega_{\new} = \frac{1}{2} \omega_{\old},
\ee
and the new solution $\vec{\bm{u}}_d^{(n+1)}$ and residual $\vec{\bm{b}}_d^{(n+1)}$ is recalculated until Eq.~\eqref{eq:convcond} holds or until $\omega < 0.001$ at which point the algorithm terminates. If the convergence condition does hold then $\vec{\bm{u}}_{d}^{(n+1)}$ is accepted and the relaxation factor grows by,
\be
\omega_{\new} = \frac{3}{2} \omega_{\old}.
\ee
Iterations are stopped when the maximum norm of the residual satisfies the condition,
\be
\label{eq:btol}
\left\Vert \bm{b}_{d} \right\Vert < \mrm{tol} = 10^{-5}.
\ee

We analytically compute both the residual and the Jacobian from our set of field equations in the symbolic manipulation software {\texttt{Maple 2018}}, which are then automatically exported into the C programming language for evaluation by our algorithm. These are then used by an iterative linear solver described in the next section.

\subsection{Linear System Solvers}
\label{ssec:LS}

The solution of linear systems of the form ${\mcl{A}} \bm{x} = \bm{b}$ such as Eq.~\eqref{eq:fullnewt} is typically done by either direct or iterative methods. Direct methods find the solution through a finite number of steps. For example, LU decomposition uses gaussian elimination to decompose the matrix ${\mcl{A}}$ into a product of lower and upper triangular matrices which simplifies the computation of the solution. Iterative methods, on the other hand, gradually approach the solution by recursively minimizing the estimated error between the current solution and the exact solution. For very large matrices, direct methods become unwieldy and can become computationally cumbersome so we use two iterative methods here.

The most common type of iterative method used today are Krylov subspace methods. These methods generate a sequence of approximate solutions from the Krylov subspace of ${\mcl{A}}$ and the residual of the linear system defined as $\bm{r} = \bm{b} - {\mcl{A}} \bm{x}$ such that the corresponding residuals converge to the zero vector. The two popular forms of this type of methods used here are the generalized minimal residual (GMRES)~\cite{doi:10.1137/0907058} and the biconjugate gradient stabilized (BiCGSTAB)~\cite{doi:10.1137/0913035} methods. The error tolerance of both iterative methods was chosen as $\mrm{LS}_{\mbox{\tiny tol}} = 10^{-12}$ which effectively places a lower bound on our numerical accuracy.

The GMRES method uses the Arnoldi method to create a Krylov subspace from the Gram-Schmidt process. The method then computes the upper triangular matrix representation of ${\mcl{A}}$ in the Krylov basis. The solution is obtained from minimizing the norm of the residual of the system in this basis. BiCGSTAB is a modified variant of the conjugate gradient method which uses a method of gradient descent to find the minimum of the system of linear equations. BiCGSTAB is the generalized version of this method that applies to non-self-adjoint matrices and uses subroutine applications of GMRES to stabalize the conjugate gradient method.

\section{Validation}
\label{sec:GR}

In this section we describe the steps we have taken to validate our algorithm. We first consider a simple ordinary differential equation. This toy problem is a useful example to demonstrate the structure of the Jacobian and to describe the detailed steps involved in applying the Newton-Raphson method.  We then move on to General Relativity and consider static, spherically symmetric vacuum solutions to the Einstein equations.  This demonstrates how the algorithm handles nonlinear coupled ordinary differential equations that represent black holes with coordinate singularities at the location of the horizon. 

\subsection{Toy Problem}

Consider a simple second order ordinary differential equation in the operator notation of Eq.~\eqref{eq:Dust}
\be
\label{eq:Dtoy}
\mcl{D} u^{\ast} = \left[\frac{d^2 }{d x^2} + 2 \frac{d }{d x} + 1 \right] u^{\ast} = 0.
\ee
The general solution to this differential equation can be found analytically to be
\be
u^{\ast}(x) = c_1 e^{-x} + c_2 x e^{-x}.
\ee
If we choose the boundary conditions, $u^{\ast}(0) = 1$ and $u^{\ast}(1) = 0$, the solution becomes
\be
u^{\ast}(x) = e^{-x} - x e^{-x}.
\ee

Let us now apply our computational infrastructure to this simple problem. The system is a single ordinary differential equation (M=1) that we wish to solve on a uniform discretized grid of $N = 101$ points. Let us choose the following initial guess for our generic vector $u$:
\be
\label{eq:toyguess}
u^{(0)}(x) = 1 - x,
\ee
where the superscript $(0)$ stands for the iteration number, i.e.~the initial guess is the $n=0$ iteration. The residual vector $b$ on the right-hand side of Eq.\eqref{eq:Diffeq} can be analytically evaluated to
\be
b^{(0)} = \mcl{D} u^{(0)} = - 1 - x,
\ee
which clearly does not vanish anywhere in the $x$ domain, i.e.~in $x \in [0,1]$. 

Since the residual does not vanish, we must correct the initial guess by some amount $\Delta u$ in the next (first) iteration. In order to find this correction, however, we must first find a discrete representation of $u(x)$ in terms of a Newton polynomial on a uniform grid at points $\{x_{i}\}$ with $i \in [0,100]$. For this toy problem, we pick a Newton polynomial of order $r = 2$, and use a centralized stencil $s_{j} = \left(I, I+1, I-1\right)$, where $I$ is the counter of the element $x_{I}$ that is closest to the value of $x$. 

With this at hand, Eq.~\eqref{eq:Newt-pol} yields
\begin{align}
\label{eq:Newt-pol-toy}
u_{d}(x) &= 
u_{s_{0}} w_{s_0,s_0} P_{s_0} (x)
+  \sum \limits_{k=0}^{1} u_{s_k} w_{s_k, s_1} P_{s_1}(x)
\nn \\
&+ \sum \limits_{k=0}^{2} u_{s_k} w_{s_k,s_2} P_{s_2} (x).
\end{align}
The basis $P_{s_{j}}(x)$ is defined in Eq.~\eqref{eq:basis}, and since the order of the polynomial here is $r=2$, the only basis that contribute are
\begin{align}
P_{s_{0}} &= 1\,, \qquad P_{s_{1}} = \left(x - x_{s_{0}}\right)\,,
\nn \\
P_{s_{2}} &= \left(x - x_{s_{0}}\right) \left(x - x_{s_{1}}\right)\,.
\end{align}
The coefficients $u_{s_{i}} = u^{(n)}(x_{s_{i}})$, while the weighting factors that contribute at this polynomial order are
\begin{align}
w_{s_{0},s_{0}} & = 1\,,
\nn \\
w_{s_{0},s_{1}} & = \left( x_{s_{0}}-x_{s_{1}}\right)^{-1}\,,
\nn \\
w_{s_{1},s_{1}} & = \left( x_{s_{1}}-x_{s_{0}}\right)^{-1}\,,
\nn \\
w_{s_{0},s_{2}} & = \left[\left( x_{s_{0}}-x_{s_{1}}\right) \left(x_{s_{0}}-x_{s_{2}} \right) \right]^{-1}\,,
\nn \\
w_{s_{1},s_{2}} & = \left[\left( x_{s_{1}}-x_{s_{0}}\right) \left(x_{s_{1}}-x_{s_{2}} \right) \right]^{-1}\,,
\nn \\
w_{s_{2},s_{2}} & = \left[\left( x_{s_{2}}-x_{s_{0}}\right) \left(x_{s_{2}}-x_{s_{1}} \right) \right]^{-1}\,.
\end{align}
Putting all of this together, we then have the discrete representation of $u(x)$ shown below in Eq.~\eqref{eq:explicit-form-toy}.

The Newton polynomial representation of order $r=2$ is the second-order, discrete Taylor expansion of the function about the point $x = x_{I}$.  In practice, the only values that $x$ can take are in the set $\{x_{i}\}$. Since this is a closed-form representation of the $u$ as a function of $x$, we can evaluate the function and all derivatives at a point $x_{i}$ straightforwardly
\bal
\bsp
u_{d}(x_{i}) &= u^{(n)}(x_{I}), \\
\pd_{x} u_{d} (x_{i}) &= \frac{u^{(n)}(x_{I+1}) - u^{(n)}(x_{I-1})}{2 \Delta x}, \\
\pd_{xx} u_{d} (x_{i}) &= \frac{u^{(n)}(x_{I+1}) - 2 u^{(n)}(x_{I}) + u^{(n)}(x_{I-1})}{\Delta x^2}, \\
\esp
\end{align}
where $\Delta x = x_{I+1}-x_{I} = x_{I}-x_{I-1}$, as must be the case since Eq.~\eqref{eq:explicit-form-toy} is a discrete Taylor expansion.  

\begin{widetext}
\begin{align}
\label{eq:explicit-form-toy}
&u_{d}(x) = 
u^{(n)}(x_{I})
\nn \\
&+  \left[ \frac{u^{(n)}(x_{I})}{x_{I}-x_{I+1}}  
+ \frac{u^{(n)}(x_{I+1})}{x_{I+1}-x_{I}}\right]  \left(x - x_{I}\right)
\nn \\
&+ \left[ \frac{u^{(n)}(x_{I})}{\left( x_{I}-x_{I+1}\right) \left(x_{I}-x_{I-1} \right)}   
+ \frac{u^{(n)}(x_{I+1})}{\left( x_{I+1}-x_{I}\right) \left(x_{I+1}-x_{I-1} \right)}  
+ \frac{u^{(n)}(x_{I-1})}{\left( x_{I-1}-x_{I}\right) \left(x_{I-1}-x_{I+1} \right)} \right]
 \left(x - x_{I}\right) \left(x - x_{I+1}\right).
\end{align} 
\end{widetext}

With this representation, we can now evaluate the discretized residual $b_{d}$, which is simply given by
\be
b_{d}(x_i) = \mcl{D} u_{d}(x_i)\,.
\ee
Using the above expressions for the analytic derivatives of the discretized function $u$, we then have that for the discretized residual at iteration $n$ 
\bal
\bsp
b_{d}^{(n)}(x_i) &=  \frac{u^{(n)}(x_{I+1}) - 2 u^{(n)}(x_{I}) + u^{(n)}(x_{I-1})}{\Delta x^2} 
\\
&+ 2 \frac{u^{(n)}(x_{I+1}) - u^{(n)}(x_{I-1})}{2 \Delta x} + u^{(n)}(x_{I}), \\
\esp
\end{align}
and at the boundaries
\bal
\bsp
b_{d}^{(n)}(x_{0}) ={}& u^{(n)}(x_{0}) - 1, \\
b_{d}^{(n)}(x_{100}) ={}& u^{(n)}(x_{100}) - 0.
\esp
\end{align}

With the discretized residual, we can now evaluate the Jacobian. From Eq.~\eqref{eq:fullnewt}, we have that 
\be
J_{i,j} = \frac{\pd b_{d}^{(n)}(x_{i})}{\pd u^{(n)}(x_{j})}
\ee
in our toy problem, whose only non-vanishing components are 
\bal
\bsp
J_{0,0} &= 1, \\
J_{I,{I-1}} &= \frac{1}{\Delta x^2} - \frac{1}{\Delta x}, \\
J_{I,I} &= -\frac{2}{\Delta x^2} + 1, \\
J_{I,{I+1}} &= \frac{1}{\Delta x^2} + \frac{1}{\Delta x}, \\
J_{{100},{100}} &= 1,
\esp
\end{align}
which is a tridiagonal matrix. Given this, the discretization error calculated from Eq.~\eqref{eq:De} is
\bal
\bsp
D_{e}(x_{i}) =& \, 2 \lb \frac{u^{(n)}(x_{I+1}) - u^{(n)}(x_{I-1})}{6 \Delta x} 
\right. 
\nn \\
& \left. - \frac{u^{(n)}(x_{I+2}) - u^{(n)}(x_{I-2})}{12 \Delta x} \rb \\
&- \frac{u^{(n)}(x_{I})}{2 \Delta x^2} + \frac{u^{(n)}(x_{I+1}) + u^{(n)}(x_{I-1})}{3 \Delta x^2} 
\nn \\
&- \frac{u^{(n)}(x_{I+2}) + u^{(n)}(x_{I-2})}{12 \Delta x^2}, \\
\esp
\end{align}
with a slightly modified formula near the boundary. 

With all of these quantities calculated, we now apply the Newton-Raphson method to invert Eq.~\eqref{eq:fullnewt}. Let us then return to vector notation and re-introduce
\be
b_{d,i}^{(n)} := b_{d}^{(n)}(x_{i})\,, \qquad
\Delta u_{d,i}^{(n)} := \Delta u_{d}^{(n)}(x_{i})\,,
\ee
such that Eq.~\eqref{eq:fullnewt} becomes
\be
J_{i,j}^{(n)} \Delta u_{d,i}^{(n)} = -b_{d,i}^{(n)}.
\ee
The inversion of this equation then yields the correction to our $n$th solution, namely $u_{d,i}^{(n+1)} = u_{d,i}^{(n)} + \Delta u_{d,i}^{(n)}$. Applying this algorithm, we find that our computational infrastructure converges to the tolerance required in a single iteration. In fact, after a single iteration, $\left\Vert {b}_{d,i} \right\Vert \approx \mcl{O} \lb 10^{-12} \rb$ even after only requiring the tolerance of Eq.~\eqref{eq:btol}. Although the algorithm efficiently minimizes the residual of the system, the residual does not directly correlate with the error between the numerical solution and the exact solution. This is due to the additional errors that are introduced during the discretization procedure described in Sec.~\ref{ssec:mindiscerror}. Therefore, the ``true" error between the numerical solution and the exact solution is determined by a combination of the discretized residual and the discretization error.

\begin{figure*}[htb]
\begin{center}
\resizebox{8cm}{!}{\include{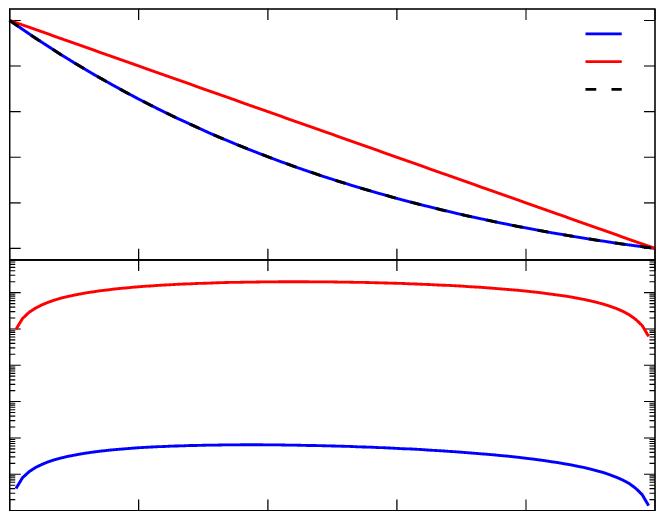}}
\resizebox{8cm}{!}{\include{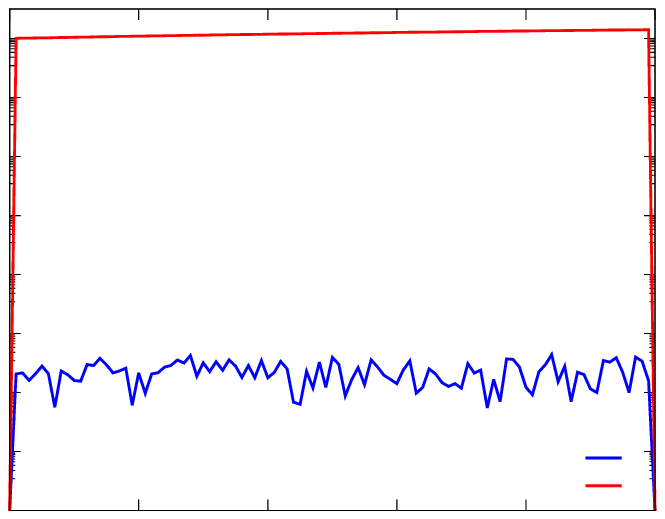}}
\caption{\label{fig:Toyprobsol} (Color online) Numerical solution to the toy problem (top left), true error between numerical and exact solution (bottom left), and residual (right). The red line, blue line, and black dashed line is the initial guess, the numerical solution after 1 iteration, and the exact solution respectively. Observe how the residual drops to $10^{-12}$ after the first iteration, even though the required tolerance was set to only $10^{-5}$. Observe also the discrepancy between the true error and the residual after 1 iteration.}
\end{center}
\end{figure*}

In this toy problem, although the residual was minimized to $\mcl{O}(10^{-12})$, the relative discretization error is still $\mcl{O}(10^{-5})$ from Eq.~\eqref{eq:adstepcond}, so the ``true" error after 1 iteration is still limited to $\mcl{O}(10^{-5})$ (see the bottom left panel of Fig.~\ref{fig:Toyprobsol}). This implies that even though the residual converges far below our desired tolerance, the ``true" error is only just below the desired tolerance because the ``true" error is also indirectly influenced by the discretization error. The solution, error, and the residual are shown in Fig.~\ref{fig:Toyprobsol}.
%

\subsection{Schwarzschild Black Hole}

In the previous section we solved a simple ordinary differential equation using the method described in Sec.~\ref{sec:NM}. We would now like to apply it to solving the elliptic differential equations that arise from the vacuum Einstein equations in spherical symmetry and stationarity. The solution is the well-known Schwarzschild metric to which we can compare our numerical results. We therefore use this example as a benchmark of the performance of our algorithm. 

The Einstein-Hilbert action in General Relativity in a vacuum is given by
\be
\label{eq:GRaction}
S = \frac{1}{16 \pi} \int d^4 x \sqrt{-g} \; R,
\ee
where $R$ is the Ricci scalar and $g$ is the determinant of the metric $\tn{g}{_\mu_\nu}$. Varying the action with respect to the metric gives the vacuum Einstein field equations
\be
\label{eq:GREE}
\tn{G}{_\mu_\nu} = 0\,,
\ee
where $G_{\mu \nu}$ is the Einstein tensor. 

Let us now consider a spherically symmetric and stationary metric ansatz in isotropic coordinates
\be
\label{eq:ansatz}
ds^2 = - f (\rho) \, dt^2 + m (\rho) \lsb d \rho^2 + \rho^2 d \Omega^2 \rsb,
\ee
where $\rho$ is the isotropic radial coordinate, which is related to the Schwarzschild radial coordinate by 
\be
r = \rho [ 1 + {r_{\Hz}}/({4 \rho})]^2\,,
\ee
where $r_{\Hz} = 2 M_0$ is the horizon radius in Schwarzschild coordinates and $M_0$ is the mass. 

With this ansatz, the coupled system of ($M=2$ in our computational infrastructure notation) differential equations that we wish to solve can be found from the $(t,t)$ and $(\rho,\rho)$ components of the Einstein tensor. These components are
\bal
\label{eq:Gtt}
\tn{G}{_t_t} &= -\frac{f \; m^{\prime \prime}}{m^2}  + \frac{3 f \lb m^{\prime}\rb^2}{4 m^3}  - \frac{2 f \; m^{\prime}}{\rho \, m^2},
\\
\tn{G}{_\rho_\rho} &= \frac{\lb m^{\prime} \rb^2}{4 m^2}  + \frac{m^{\prime} f^{\prime}}{2 f m} + \frac{m^{\prime}}{\rho \, m}  + \frac{f^{\prime}}{\rho \, f},
\label{eq:Grhorho}
\end{align}
with primes standing for radial derivatives. 

The Schwarzschild solution to these equations in these coordinates is
\bal
\bsp
\label{eq:Schw-metric}
f_{\GR} = \lb \frac{1 - \frac{r_{\Hz}}{4 \rho}}{1 + \frac{r_{\Hz}}{4 \rho}} \rb^2, \qquad
m_{\GR} = \lb 1 + \frac{r_{\Hz}}{4 \rho} \rb^4.
\esp
\end{align}
The event horizon in isotropic coordinates is located at $\rho = \rho_{\Hz} = r_{\Hz}/4$, as found from the condition $\tn{g}{_t_t}|_{\rho=\rho_{\Hz}} = f_{\GR} |_{\rho=\rho_{\Hz}} = 0$. 

The boundary conditions at the event horizon and at spatial infinity are determined by regularity and smoothness. At the event horizon we must have
\be
f_{\GR} |_{\rho=\rho_{\Hz}} = 0, \qquad m_{\GR} |_{\rho=\rho_{\Hz}} = 16,
\ee
which follows from evaluation of the analytic solution at $\rho = \rho_{\Hz}$. At spatial infinity, asymptotic flatness requires that
\be
f_{\GR} |_{\rho \rarr \infty} = 1, \qquad m_{\GR} |_{\rho \rarr \infty} = 1.
\ee

Our computational infrastructure allows us to not only find the numerical solution to Eq.~\eqref{eq:GREE}, but also to find some observable global quantities that characterize the black hole spacetime. Asymptotically near spatial infinity, the leading order terms of the fields decay as 
\bal
\begin{split}
\label{eq:asymr}
f_{\GR} &= 1 - \frac{2 M}{\rho} + \mcl{O} \lb \frac{1}{\rho^2} \rb, \\
m_{\GR} &= 1 + \frac{2 M}{\rho} + \mcl{O} \lb \frac{1}{\rho^2} \rb, \\
\end{split}
\end{align}
where $M$ is the Arnowit-Deser-Misner (ADM) mass. Therefore, the ADM mass can be found from the coefficient of the $1/\rho$ expansion of the numerical solution near spatial infinity. Because of the high Newton polynomial order we use, we can interpolate our numerical solutions very close to spatial infinity to a high degree of accuracy, which allows us to extract the ADM mass from our numerical solutions easily.  

The extraction of the ADM mass and the imposition of boundary conditions becomes more precise through the use of a compactified coordinate. We therefore introduce the coordinate $x$, defined by
\be
\label{eq:xcomp}
x = \frac{1 - r_{\Hz}/ 4 \rho}{1 + r_{\Hz} /4 \rho},
\ee
and perform a coordinate transformation prior to solving our differential system. This changes our domain of integration from $\rho~\in~[r_{\Hz}/4,\infty)$ to the finite domain $x~\in~[0,1]$. In these compactified isotropic coordinates, the Schwarzschild solution has the form
\be
f_{\GR} = x^2, \qquad m_{\GR} = \frac{16}{\lb 1 + x \rb^4}.
\ee

With the differential system in compactified coordinates, we then solve the problem numerically as specified in Sec.~\ref{sec:NM}. We begin by replacing each function and differential operator with a discretized Newton polynomial representation of order $r = 12$ on a grid of $N = 101$ points. We then initialize the numerical solver with a initial guess that is a small perturbation away from the Schwarzschild metric and that vanishes at the boundaries\footnote{For this toy problem, we know the exact analytic solution \emph{a priori}, so we could initialize our solver with it. Doing so, however, would prevent us from validating our computational infrastructure.}
\bal
\begin{split}
u_{0}^{(0)} ={}& f_{\GR} \left[ 1 + \delta \; x \; \lb 1 - x \rb \right], \\
u_{1}^{(0)} ={}& m_{\GR} \left[ 1 + \delta \; x \; \lb 1 - x \rb \right],
\end{split}
\end{align}
where $\delta = 0.1$. One can adjust $\delta$ to improve or worsen the initial guess, which in turn affects the number of iterations required to converge to a solution within the tolerance required. 

Applying this algorithm, we find that our computational infrastructure converges to the desired tolerance in 3 iterations. The number of iterations is related to the initial guess, which in this case is controlled by the value of $\delta$. In the limit as $\delta \rarr 0$ the initial guess becomes the exact solution, and the initial residual decreases below tolerance to within numerical precision.

\begin{figure*}[htb]
\begin{center}
\resizebox{8cm}{!}{\include{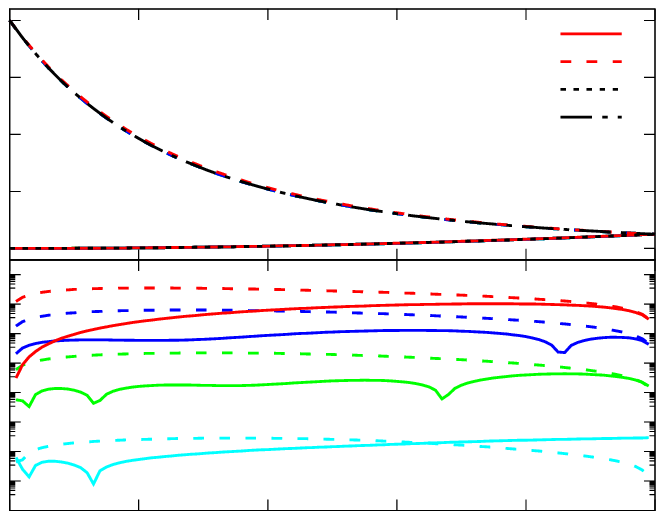}}
\resizebox{8cm}{!}{\include{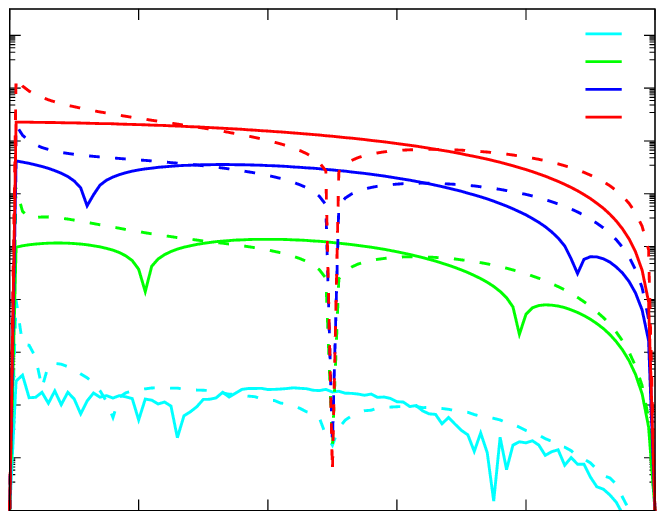}}
\caption{\label{fig:schwsol} (Color online) Numerical solution (top left), error between the numerical solution and the exact solution (bottom left), and residual (right) per iteration for the vacuum Einstein equations in spherical symmetry. The solid and dashed lines indicate the $\tn{g}{_t_t}$ and $\tn{g}{_\rho_\rho}$ component of the metric respectively. Different colors indicate iteration number. Observe that the numerical solution converges to the Schwarzschild metric within three iterations and that the residual does closely mirror the error in this example.}
\end{center}
\end{figure*}

Unlike in the toy problem from the previous section, the ``true" error between the numerical solution and the exact solution is now of comparable order to the residual $\mcl{O}(10^{-12})$. This is due to the closed polynomial form of the Schwarzschild solution in compactified isotropic coordinates, which are very well approximated by our Newton polynomial. The comparison between the toy problem and the Schwarzschild application suggests that in problems where we do not have the exact solution to compute the ``true" error we cannot use the residual as a direct measure of the error. In other words, even if we have a minimized residual far below the desired tolerance, the solution must be assumed to only be accurate to the desired tolerance. The numerical solution and residuals are shown in Fig.~\ref{fig:schwsol}.
%

\section{Spherically Symmetric Black Holes in Scalar-Gauss-Bonnet Gravity}
\label{sec:EdGB}

In this section we solve the modified Einstein field equations in sGB gravity with both a linear coupling and an exponential coupling function, assuming a vacuum spacetime that is also spherical symmetric and stationary.

\subsection{Action and Field equations}

The action in scalar-Gauss-Bonnet gravity in a vacuum is given by
\be
\label{eq:action}
S = \frac{1}{16 \pi} \int d^4 x \sqrt{-g} \lsb R - \beta \, \tn{\nabla}{_\mu} \psi \tn{\nabla}{^\mu} \psi + 2 \, \alpha F(\psi) \mcl{G} \rsb,
\ee
where $R$ is the Ricci scalar and $g$ is the determinant of the metric $\tn{g}{_\mu_\nu}$. The real dimensionless scalar field $\psi$ is coupled through a coupling constant $\alpha$ which has dimensions of length squared  and a function of the scalar field $F(\psi)$ to the Gauss-Bonnet invariant~$\mcl{G}$
\be
\mcl{G} = R^2 - 4 \tn{R}{^\mu^\nu} \tn{R}{_\mu_\nu} + \tn{R}{^\mu^\nu^\rho^\sigma} \tn{R}{_\mu_\nu_\rho_\sigma}.
\ee
We keep the coupling constant $\beta$ around in this section, but in all computation we set $\beta = 1$, as it can be eliminated through a redefinition of the scalar field $\psi$ and the coupling constant $\alpha$. Note this form of the action differs from that introduced in~\cite{Yunes2013} by a factor of $\kappa = (16 \pi)^{-1}$ such that $\tilde{\beta} = 2\, \kappa \beta$ and $\tilde{\alpha} = 2\, \kappa \alpha$.

By varying the action with respect to the metric and the scalar field we obtain two field equations. Variation with respect to the metric field yields
\be
\label{eq:EE}
\tn{G}{_\mu_\nu} - \beta \, \tn{T}{_\mu_\nu} + \alpha \, \tn{K}{_\mu_\nu} = 0,
\ee
where the scalar field stress-energy tensor is
\be
\label{eq:T}
\tn{T}{_\mu_\nu} = \tn{\cd}{_\mu} \psi \tn{\cd}{_\nu} \psi - \frac{1}{2} \tn{g}{_\mu_\nu} \tn{\cd}{^\gamma} \psi \tn{\cd}{_\gamma} \psi,
\ee
and
\bal
\begin{split}
\label{eq:Kdef}
\tn{K}{_\mu_\nu} &= \lb \tn{g}{_\rho_\mu} \tn{g}{_\delta_\nu} + \tn{g}{_\rho_\nu} \tn{g}{_\delta_\mu} \rb \times \\
& \tn{\cd}{_\sigma} \lsb \tn{\epsilon}{^\gamma^\delta^\alpha^\beta} \tn{\epsilon}{^\rho^\sigma^\lambda^\eta} \tn{R}{_\lambda_\eta_\alpha_\beta} \tn{\cd}{_\gamma} F(\psi) \rsb.
\end{split}
\end{align}
Variation with respect to the scalar field yields
\be
\label{eq:box}
\beta \Box \psi + \alpha \, \frac{\pd F}{\pd \psi} \mcl{G} = 0.
\ee
The scalar field is subject to the following boundary conditions:  it must be asymptotically flat, and its first derivative must vanish on the horizon in isotropic coordinates,\footnote{This follows from the regularity condition on the horizon~\cite{PhysRevD.54.5049, Sotiriou:2013qea,PhysRevD.90.124063}.} namely
\be
\label{eq:psiBC}
\frac{\pd \psi}{\pd \rho} |_{\rho \rarr \rho_{\Hz}} = 0, \qquad \psi |_{\rho \rarr \infty} = 0.
\ee

Utilizing the spherically symmetric metric ansatz in isotropic coordinates from before, the Einstein tensor is still given by Eqs.~\eqref{eq:Gtt} and~\eqref{eq:Grhorho}, and the two new sets of terms in the field equations are:
\bal
\begin{split}
\tn{T}{_t_t} &= \frac{f}{2 m} \lb \psi^{\prime} \rb^2, \qquad
\tn{T}{_\rho_\rho} = \frac{1}{2} \lb \psi^{\prime} \rb^2, \\
\tn{T}{_\theta_\theta} &= - \frac{\rho^2}{2} \lb \psi^{\prime} \rb^2, \qquad
\tn{T}{_\phi_\phi} = \ssqth \, \tn{T}{_\theta_\theta}, \\
\end{split}
\end{align}
where the primes denote derivatives with respect to the $\rho$ coordinate, e.g.~$f^{\prime} = \frac{d f}{d \rho}$ and $\tn{K}{_\mu_\nu}$ is given below in Eq.~\eqref{eq:K}.

Let us end this subsection with a discussion of the coupling function $F \lb \psi \rb$. In full EdGB gravity, the coupling function is $F \lb \psi \rb = \exp({\psi})$, but in the regime where $\psi$ is small, we can Taylor expand the coupling function as $F \lb \psi \rb = 1 + \psi + \mcl{O}(\psi^2)$. The $\psi$-independent term in this expansion is irrelevant as it leads to a theory that is identical to GR due to the Gauss-Bonnet invariant being a topological invariant. In this paper, we focus on numerical calculations for sGB gravity with both an exponential coupling and a linear coupling, namely for theories defined by  
\begin{align}
\begin{split}
F(\psi) &= \psi \; \; \, \leftrightarrow \;  {\rm{linear}} \; {\rm{sGB}}\,,
\\
F(\psi) &= e^{\psi} \; \leftrightarrow \; {\rm{EdGB}}\,.
\end{split}
\end{align}
We consider each of these cases separately in the next subsections. 
\bw
\bal
\begin{split} \label{eq:K}
\tn{K}{_t_t} &= F^{\prime \prime} \lsb \frac{2 f}{m^4} \lb m^{\prime} \rb^2 + \frac{8 f}{m^3 \rho} \lb m^{\prime} \rb \rsb \qquad \tn{K}{_\phi_\phi} ={} \ssqth \, \tn{K}{_\theta_\theta}, \\
&+ F^{\prime} \lsb \frac{4 f}{m^4} \lb m^{\prime \prime} \rb \lb m^{\prime} \rb - \frac{5 f}{m^5} \lb m^{\prime} \rb^3 + \frac{8 f}{m^3 \rho} \lb m^{\prime \prime} \rb - \frac{8 f}{m^4 \rho} \lb m^{\prime} \rb^2 + \frac{8 f}{m^3 \rho^2} \lb m^{\prime} \rb \rsb, \\
\tn{K}{_\rho_\rho} &= F^{\prime} \lsb - \frac{3}{f \, m^3} \lb f^{\prime} \rb \lb m^{\prime} \rb^2 - \frac{12}{f \, m^2 \rho} \lb f^{\prime} \rb \lb m^{\prime} \rb - \frac{8}{f \, m \rho^2} \lb f^{\prime} \rb \rsb, \\
\tn{K}{_\theta_\theta} &= F^{\prime \prime} \lsb - \frac{2 \rho^2}{f \, m^2} \lb f^{\prime} \rb \lb m^{\prime} \rb - \frac{4 \rho}{f \, m} \lb f^{\prime} \rb \rsb \\
&+ F^{\prime} \bigg[ - \frac{2 \rho^2}{f \, m^2} \lb m^{\prime \prime} \rb + \frac{4 \rho^2}{f \, m^3} \lb f^{\prime} \rb \lb m^{\prime} \rb^2 - \frac{2 \rho^2}{f \, m^2} \lb f^{\prime \prime} \rb \lb m^{\prime} \rb + \frac{\rho^2}{f^2 m^2} \lb f^{\prime} \rb^2 \lb m^{\prime} \rb \\
&+ \frac{2 \rho}{f \, m^2} \lb f^{\prime} \rb \lb m^{\prime} \rb - \frac{4 \rho}{f \, m} \lb f^{\prime \prime} \rb + \frac{2 \rho}{f^2 m} \lb f^{\prime} \rb^2 \bigg]. \\
\end{split}
\end{align}
\ew
%

\subsection{Linear Scalar-Gauss-Bonnet Gravity}

In the linear coupling theory, we can find an analytical perturbative solution to the field equations assuming the small-coupling limit, i.e.~$\bar{\alpha} := \alpha / r_{\Hz}^2 \ll 1$ because $r_{\Hz}$ is of the order of the curvature length of the system under consideration and we introduce the dimensionless coupling constant $\bar{\alpha}$. Let us then use a deformed-Schwarzschild ansatz for the metric tensor
\bal
\begin{split}
ds^2 ={}& - \lb f_0 + \epsilon f_1 + \epsilon^2 f_2 \rb dt^2 \\
{}& + \lb m_0 + \epsilon m_1 + \epsilon^2 m_2 \rb \lsb d \rho^2 + \rho^2 d \Omega^2 \rsb,
\end{split}
\end{align}
where $\epsilon \ll 1$ is a book-keeping parameter and $\bar{\alpha}$ is $\mcl{O}(\epsilon)$, and the following ansatz for the scalar field
\be
\psi = \psi_0 + \epsilon \psi_1 + \epsilon^2 \psi_2.
\ee
Both of these ansatz are assumed valid up to ${\mcl{O}}(\epsilon^{3})$. 

Inserting the ansatz in the field equations, we can analytically solve for the metric and the scalar field order by order in $\epsilon$, imposing regularity on the horizon and asymptotic flatness at spatial infinity to fix any integration constants. At $\mcl{O}(\epsilon^{0})$, $f_{0}$ and $m_{0}$ are just the Schwarzschild metric in isotropic coordinates of Eq.~\eqref{eq:Schw-metric}, while the scalar field vanishes $\psi_{0} = 0$ due to asymptotic flatness. At $\mcl{O}(\epsilon)$, we find the metric perturbations vanish, while the scalar field perturbation is
\bal
\begin{split}
\label{eq:psi1}
f_1 ={}& 0, \qquad
m_1 = 0, \\
\psi_1 ={}& \frac{4 \bar{\alpha}}{\beta \rho \, r_{\Hz}^2} \frac{\lb 1 + \frac{3 r_{\Hz}}{2 \rho} + \frac{23 r_{\Hz}^2}{24 \rho^2} + \frac{3 r_{\Hz}^3}{32 \rho^3} + \frac{r_{\Hz}^4}{256 \rho^4} \rb}{\lb 1 + r_{\Hz}/4 \rho \rb^6}.
\end{split}
\end{align}

At $\mcl{O}(\epsilon^2)$, we find the first nontrivial correction to the metric tensor and we find that the scalar field perturbation at this order vanishes. Both are given below in isotropic coordinates in Eq.~\eqref{eq:f2m2}. We can then further express these analytic solutions in compactified coordinates for later use which are shown in Eq.~\eqref{eq:linsolx}. These results agree exactly with those found in~\cite{PhysRevD.83.104002} and~\cite{PhysRevD.90.124063} after performing the coordinate transformation from Schwarzschild to isotropic and compactified isotropic coordinates. 
\bw
\bal
\label{eq:f2m2}
\begin{split}
f_2(\rho) ={}& \frac{\bar{\alpha}^2}{\beta \lb 1 + r_{\Hz}/ 4 \rho \rb^{14}} \bigg[ -\frac{49 r_{\Hz}}{5 \rho} - \frac{49 r_{\Hz}^2}{2 \rho^2} - \frac{1637 r_{\Hz}^3}{60 \rho^3} - \frac{929 r_{\Hz}^4}{96 \rho^4} + \frac{16753 r_{\Hz}^5}{3840 \rho^5} + \frac{18893 r_{\Hz}^6}{3840 \rho^6} - \frac{5573 r_{\Hz}^7}{1920 \rho^7} \\
&+ \frac{146549 r_{\Hz}^8}{430080 \rho^8} + \frac{188761 r_{\Hz}^9}{6881280 \rho^9} - \frac{579 r_{\Hz}^{10}}{917504 \rho^{10}} - \frac{2231 r_{\Hz}^{11}}{9175040 \rho^{11}} - \frac{6553 r_{\Hz}^{12}}{48442112 \rho^{12}} - \frac{6553 r_{\Hz}^{13}}{19377684480 \rho^{13}} \bigg] ,\\
m_2(\rho) ={}& \frac{\bar{\alpha}^2}{\beta \lb 1 + r_{\Hz}/ 4 \rho \rb^8} \bigg[ \frac{49 r_{\Hz}}{5 \rho} + \frac{499 r_{\Hz}^2}{20 \rho^2} + \frac{1345 r_{\Hz}^3}{48 \rho^3} + \frac{889 r_{\Hz}^4}{64 \rho^4} - \frac{11 r_{\Hz}^5}{640 \rho^5} - \frac{28787 r_{\Hz}^6}{7680 \rho^6} + \frac{4689 r_{\Hz}^7}{71680 \rho^7} \\
&+ \frac{4303 r_{\Hz}^8}{57344 \rho^8} + \frac{4727 r_{\Hz}^9}{458752 \rho^9} + \frac{20011 r_{\Hz}^{10}}{27525120 \rho^{10}} + \frac{35149 r_{\Hz}^{11}}{1211105280 \rho^{11}} + \frac{2383 r_{\Hz}^{12}}{484421120 \rho^{12}} \bigg], \\
\psi_2(\rho) ={}& 0,
\end{split}
\end{align}
\bal
\label{eq:linsolx}
\begin{split}
f_2(x) ={}& \frac{\bar{\alpha}^2}{\beta} \frac{x \lb -1 + x^2 \rb}{1155} \lb 2383 + 154 x + 55594 x^3 - 102410 x^5 + 83094 x^7 - 32956 x^9 + 5460 x^{11} \rb, \\
m_2(x) ={}& \frac{\bar{\alpha}^2}{\beta} \frac{32 x \lb 1 - x  \rb}{1155 \lb 1+x \rb^4} ( -2383 + 25337 x + 25337 x^2 - 37033 x^3 - 37033 x^4 +, \\
{}&+ 27031 x^5 + 27031 x^6 - 10094 x^7 - 10094 x^8 + 1610 x^9 + 1610 x^{10} ), \\
\psi_1(x) ={}& \frac{\bar{\alpha}}{\beta} \frac{2 \lb 1 - x^2 \rb}{3} \lb 11 - 7 x^2 + 2 x^4 \rb.
\end{split}
\end{align}
\ew

From these solutions we can obtain the ADM mass and scalar charge from an expansion as $\rho \rarr \infty$, namely
\bal
\begin{split}
M &= \frac{r_{\Hz}}{2} \lb 1 + \frac{49}{5} \frac{\bar{\alpha}^2}{\beta} \rb + \mcl{O} (\alpha^3), \\
D &= \frac{4 \bar{\alpha}}{\beta} + \mcl{O}(\alpha^3),
\end{split}
\end{align}
where $r_{\Hz} = 2 \, M_0$ and $M_0$ is the bare mass of the black hole that appears in the background (Schwarzschild) metric. Just as the ADM mass $M$ is related to the coefficient of the $1/\rho$ term in an expansion of the metric about spatial infinity, the charge $D$ is related to the same coefficient but in the expansion of the scalar field about spatial infinity. 

\begin{figure}[h]
\begin{center}
\resizebox{8cm}{!}{\include{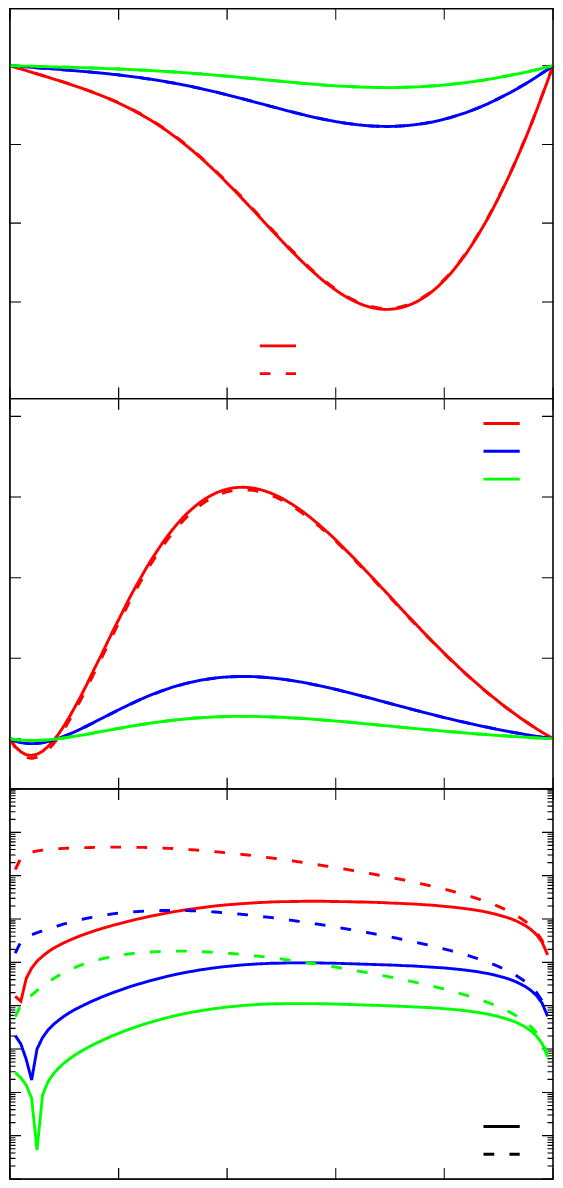}}
\caption{\label{fig:LinCgtt} Linear scalarl-Gauss-Bonnet correction to $\tn{g}{_t_t}$ (top) and $\tn{g}{_\rho _\rho}$ (middle), and difference (bottom) between the numerical and the analytic perturbative solutions as a function of the compactified coordinate $x$ and for different $\bar{\alpha}$ indicated by color. In the top two panels, the solid lines indicate the $\tn{g}{_t_t}$ or $\tn{g}{_\rho_\rho}$ component numerical solution and the dashed lines indicate the analytic perturbative solution. Conversely, in the bottom panel, the solid lines indicate the $\tn{g}{_t_t}$ and dashed lines indicate the $\tn{g}{_\rho_\rho}$ component difference with the same color scheme in the top two panels. The analytic perturbative solution agrees very well with the full numerical solution, with differences that grow only to ${\mcl{O}}(10^{-5})$.}
\end{center}
\end{figure}

With this analysis in hand, let us now focus on numerically solving the field equations, Eqs.~\eqref{eq:EE} and~\eqref{eq:box}, simultaneously for both metric functions $\tn{g}{_t_t}$ and $\tn{g}{_\rho_\rho}$ and the scalar field $\psi$ without any approximations. In order to do so, we employ the computational infrastructure described in detail in Sec.~\ref{sec:NM}. In particular, we choose an initial grid of $N = 101$ points and a Newton polynomial order $r = 12$. For the actual computation, we set $r_{\Hz} = 1$ which sets the bare mass of the black hole to $M_{0} = 1/2$. Different black hole masses can be obtained by scaling the radial coordinate appropriately. Note this renders our equations dimensionless and the correct units can be restored through a similar rescaling. The desired tolerance of the solution is $\mrm{tol} = 10^{-5}$ which is both placed on the residual and on the relative tolerance of the discretization correction in Eq.~\eqref{eq:adstepcond}. The tolerance in the iterative linear solvers described in Sec.~\ref{ssec:LS} is $\mrm{LS}_{\mbox{\tiny tol}} = 10^{-12}$ and places a lower bound on our numerical accuracy. We will also employ the compactified coordinate system defined in Eq.~\eqref{eq:xcomp} to change our domain of integration to the finite domain $x \in [0,1]$. In these compactified isotropic coordinates, the perturbative corrections of Eq.\eqref{eq:f2m2} are shown below in Eq.~\eqref{eq:linsolx}.

We can now compare the analytic perturbative solutions to the full non-linear solutions. The top and middle panels of Fig.~\ref{fig:LinCgtt} show the sGB corrections to the metric components as a function of the compactified coordinate for different choices of the coupling constant $\bar{\alpha}$. Observe that the numerical solution is almost indistinguishable from the analytic perturbative solution everywhere in the domain. The agreement is so remarkable that it is worthwhile exploring the difference between the analytic and the numerical solution, which we do in the bottom panel of Fig.~\ref{fig:LinCgtt}. Observe that the difference is indeed very small, ranging from $\mcl{O}(10^{-10})$ to almost $\mcl{O}(10^{-4})$ depending on the metric component one studies, and it increases as the coupling strength is increased as expected. We have verified that the residual is always orders of magnitude smaller than this difference in our numerical solutions. 

\begin{figure}[htb]
\begin{center}
\resizebox{8cm}{!}{\include{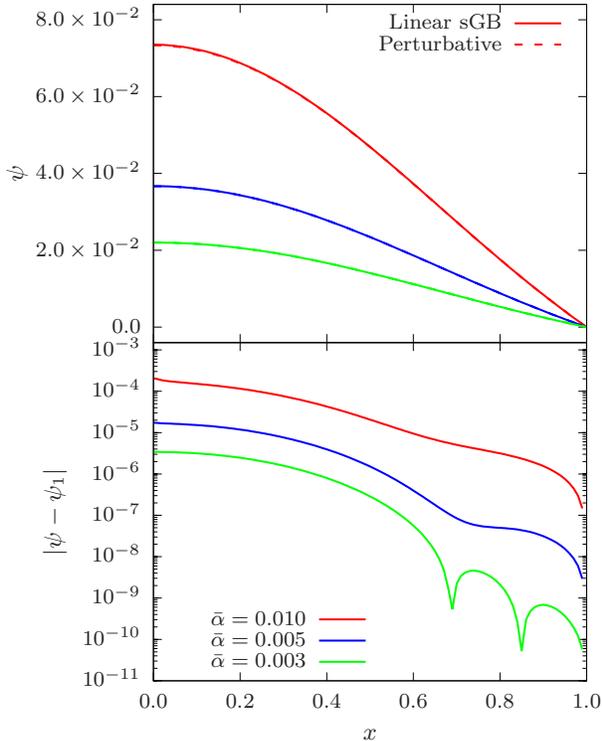}}
\caption{\label{fig:LinCpsi} Scalar field $\psi$ (top) and difference between the numerical and the analytic perturbative solution (bottom) as a function of the compactified coordinate $x$ in linear sGB. Solid lines indicate the numerical solution with linear coupling and dashed lines indicate the analytic perturbative solution, color coded for different coupling strengths $\bar{\alpha}$.}
\end{center}
\end{figure}

We observe similar behavior in the scalar field. The top panel of Fig.~\ref{fig:LinCpsi} shows the scalar field solved numerically and the analytic perturbative solution both as a function of the compactified $x$ coordinate. Observe again that the curves are right on top of each other. The difference between these curves is shown in the bottom panel of this figure, where we see clearly that the difference ranges from ${\mcl{O}}(10^{-10})$ to ${\mcl{O}}(10^{-4})$ for the largest couplings we considered. As before, we have verified that the residual of the scalar field equation of motion is smaller than this difference for all cases considered. Observe, however, that this time the difference between the numerical solution and the analytic perturbative solution is much larger than it was for either metric component. This makes sense because the modified field equations depend on the scalar field through its stress-energy tensor, which is quadratic in the scalar field. Thus, we expect a difference of ${\mcl{O}}(10^{-a})$ for some $a \in \mathbb{R}_{>0}$ to contribute a difference of ${\mcl{O}}(10^{-2a})$ to the metric components.

From these comparisons we can extract a few useful conclusions. Perhaps most importantly, we see that the non-linear corrections to the solution are truly small everywhere in the domain. This is sensible because the scalar field itself is small and the values of $\bar{\alpha}$ that we explore are small, so the corrections to GR can be treated perturbatively. Another interesting observation is that the largest deviations from GR manifest somewhere in the middle of the domain. This is in part due to the boundary conditions: at spatial infinity the metric must be asymptotically flat so the GR deformation must vanish at a suitable fall-off rate; near the horizon, the GR deformation must be regular, but due to its asymptotic form near the horizon, it must also vanish there. Our choice of quantities to compare also plays an important role. Note that we compare metric coefficients at different physical locations, as $x=0$ at the horizon by definition and the horizon can be at a different location in the two metrics we are comparing. It is also worth pointing out that observables may depend on derivatives of the metric or even integrals of combinations of metric functions in the spatial domain and, hence, agreement between metric coefficients in part of the domain does not necessarily imply that observables in linear sGB will be close to those in GR. We have seen this already in Fig.~\ref{fig:Masscomp}, and will return to this point in Sec.~\ref{sec:props}.

\subsection{Einstein-dilaton-Gauss-Bonnet Gravity}

Let us now consider the case of an exponential coupling function. The resulting field equations are Eqs.~\eqref{eq:EE} and~\eqref{eq:box} with $F(\psi) = e^{\psi}$. In this case, a perturbative solution in small coupling $\bar{\alpha}$ does not exist, but we can still compare any numerical solutions to the small coupling approximation for the linear theory of the previous section. We find a numerical solution using the computational infrastructure of Sec.~\ref{sec:NM}, with the same choices for the grid spacing, Newton polynomial order, etc as in the previous subsection.

\begin{figure}[htb]
\begin{center}
\resizebox{8cm}{!}{\include{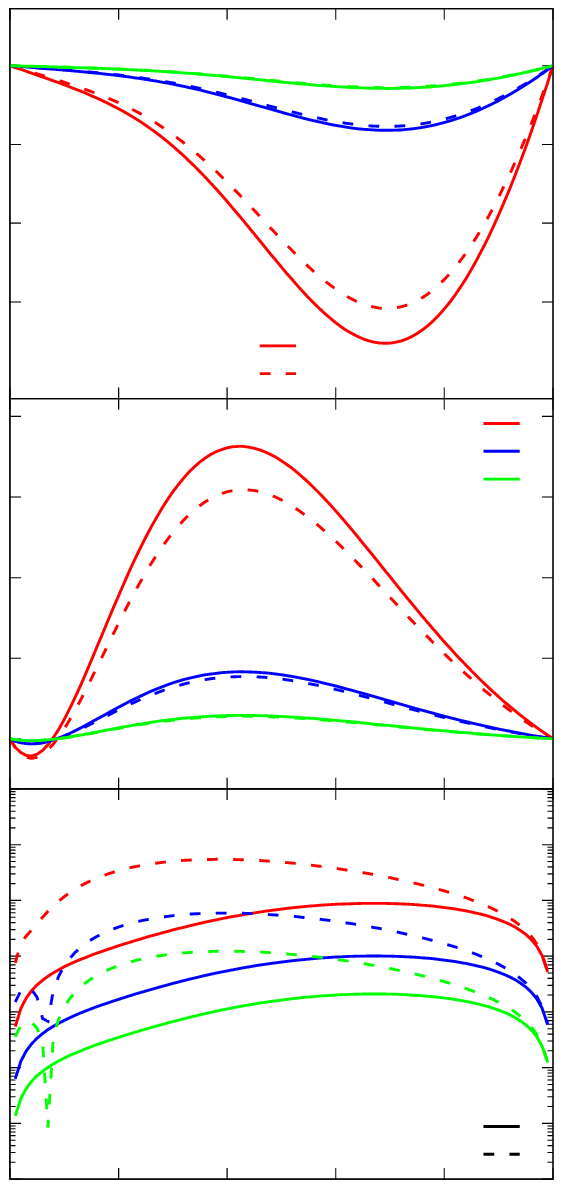}}
\caption{\label{fig:ExpCgtt} Same as Fig.~\ref{fig:LinCgtt} but for Einstein-dilaton-Gauss-Bonnet. In contrast to the differences between the linear and perturbative solution in Fig.~\ref{fig:LinCgtt}, the differences between the numerical solution in the exponential coupling case and perturbative analytic solution in the linear coupling case are much more prominent, growing to ${\mcl{O}}(10^{-5})$ for the highest couplings explored.}
\end{center}
\end{figure}

\begin{figure}[htb]
\begin{center}
\resizebox{8cm}{!}{\include{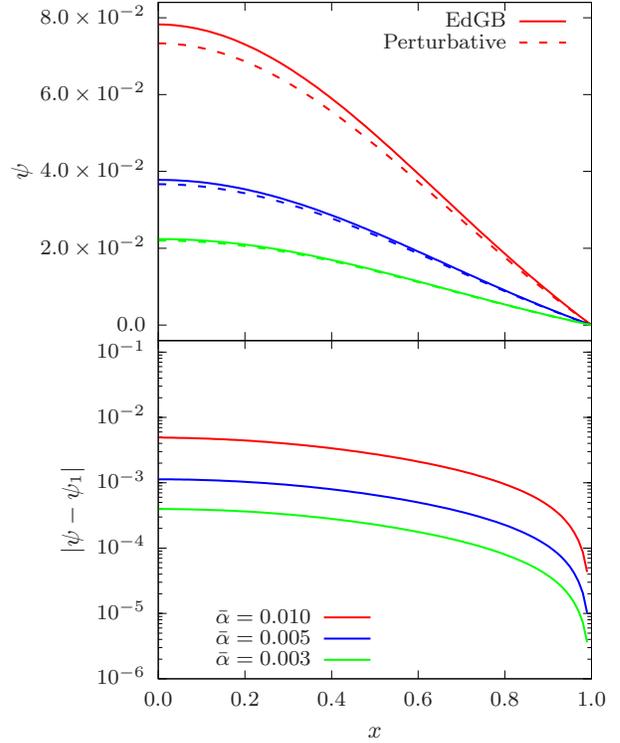}}
\caption{\label{fig:ExpCpsi} Same as Fig.~\ref{fig:LinCpsi} but for EdGB. Observe that this time the numerical solution of the exponential coupling theory differs significantly from the analytic perturbative solution of the linear coupling theory.}
\end{center}
\end{figure}

The top and middle panels of Fig.~\ref{fig:ExpCgtt} show the EdGB corrections to the metric components as a function of the compactified coordinate $x$ for different choices of the coupling constant $\bar{\alpha}$. In contrast to the linear sGB results of Fig.~\ref{fig:LinCgtt}, in EdGB the corrections to the metric are immediately noticeable. When comparing the difference between the analytic and numerical solutions in the bottom panel of Fig.~\ref{fig:LinCgtt}, we see they range from $10^{-5}$ to $10^{-3}$, whereas the difference in the linear coupling case ranged from $10^{-10}$ to $10^{-4}$. The results of this comparison are not very surprising because the analytic perturbative solution was found exclusively in the linear coupling case.
These same features can also be seen in the scalar field solution, as shown on the left panel of Fig.~\ref{fig:ExpCpsi}. The differences, shown on the right panel of this figure, range from $\mcl{O}(10^{-5})$ to $\mcl{O}(10^{-2})$, while for comparison the differences in the linear coupling case ranged from $10^{-10}$ to $10^{-4}$.

From these comparisons we can also extract a few useful conclusions. Unlike in the linear sGB case, we see here that the analytic perturbative solution found in that case does not agree with the fully nonlinear solution. This is expected. Perturbing in the coupling constant and perturbing in the scalar will formally yield the same field equations. This can be understood by the fact that one can always absorb the coupling constant in a scalar field redefinition. The range of validity of the two expansions, and hence their physical interpretation,  can be different. Nonetheless, this argument clearly implies that the perturbative solution will provide a better approximation to the linear coupling than the exponential one. Finally, as in the linear sGB case, we find that the metric corrections vanish near the horizon and compactified infinity, and both deviations asymptote to each other near infinity, but derivatives of the metric potentials can be large.

\section{Properties of Solution}
\label{sec:props}

In this section we explore the physical properties of the numerical solutions found in the previous section. We begin by finding analytical models that we fit to the data to provide accurate, closed-form expressions that allow for the rapid computation of physical observables. We then use these fitted models and the numerical results to calculate the location of the innermost stable circular orbit and the light ring, and compare them with each other and with the analytical perturbative solutions.

\subsection{Fitting Function}

In the compactified coordinate system introduced in Eq.~\eqref{eq:xcomp}, the full nonlinear solutions can be expressed as
\bal
\begin{split}
f (x) ={}& f_{\GR} + f_2 (x) + f_{\nonlin}(x), \\
m (x) ={}& m_{\GR} + m_2 (x) + m_{\nonlin}(x), \\
\psi (x) ={}& \psi_1 (x) + \psi_{\nonlin}(x), \\
\end{split}
\end{align}
where $f_2(x)$, $m_2(x)$, and $\psi_1(x)$ are the analytical perturbed solutions of Eqs.~\eqref{eq:linsolx} in the compactified coordinates and $f_{\nonlin}(x)$, $m_{\nonlin}(x)$, $\psi_{\nonlin}(x)$ are nonlinear corrections that we wish to find. 

Using the analytical perturbed solutions as an ansatz, we propose best fit models for the non-linear corrections of the form
\bal
\label{eq:nonfit}
\begin{split}
f_{\nonlin}(x) ={}& x \lb -1 + x^2 \rb \lb \sum_i \sum_j a_{i,j} \bar{\alpha}^i x^j \rb, \\
m_{\nonlin}(x) ={}& \frac{x \lb 1 - x \rb}{\lb 1 + x \rb^4} \lb \sum_i \sum_j b_{i,j} \bar{\alpha}^i x^j \rb, \\
\psi_{\nonlin}(x) ={}& \lb 1 - x^2 \rb \lb \sum_i \sum_j c_{i,j} \bar{\alpha}^i x^j \rb,
\end{split}
\end{align}
where we have set $\beta = 1$. We then fit these models to our numerical solutions to determine the constants $(a_{i,j},b_{i,j},c_{i,j})$ on the grid domain $x \in [0,1]$ and $\bar{\alpha} \in [0.0001,0.013]$. For $\bar{\alpha} < 0.0001$ the analytical perturbative solution is indistinguishable from the full nonlinear solution to our specified tolerance. For $\bar{\alpha} > 0.013$ we find pathologies in the numerical solution that we will describe in Sec.~\ref{ssec:naked}. The fitting order of our models is determined by systematically increasing the polynomial order of each function until the residual between the numerical solution and the model saturates. Some of the best-fit coefficients $(a_{i,j}, b_{k,l}, c_{m,n})$ are included in Appendix~\ref{app:coeff}, but they are available in a {\texttt{Mathematica}} file upon request. 

\begin{figure*}[htb]
\begin{center}
\resizebox{7cm}{!}{\include{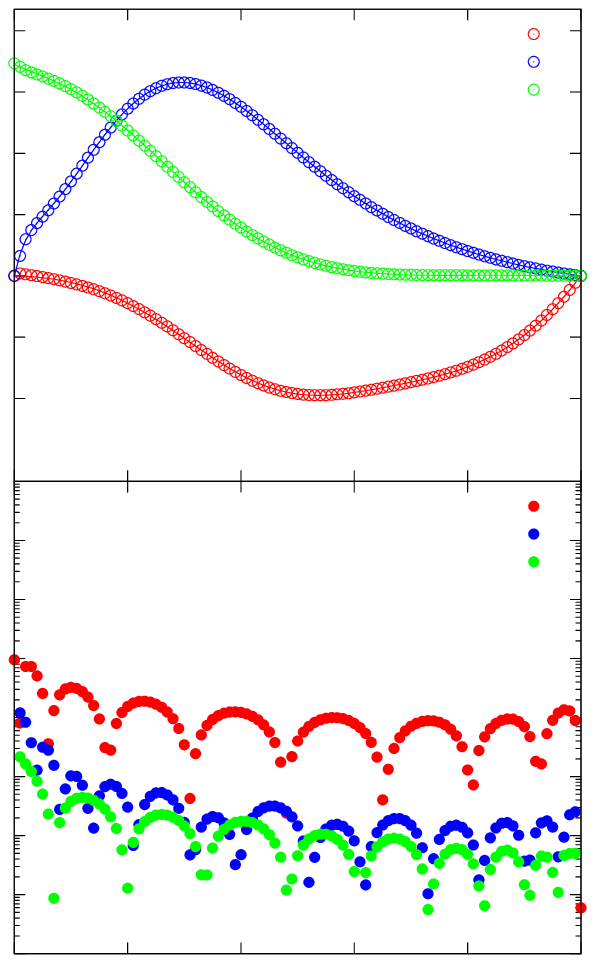}}
\resizebox{7cm}{!}{\include{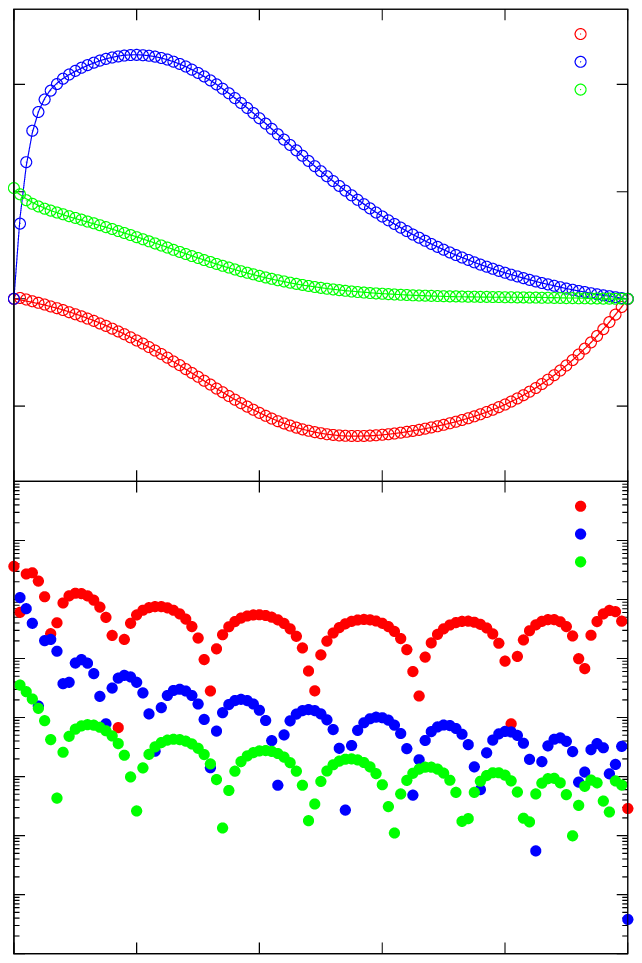}}
\\
\resizebox{7cm}{!}{\include{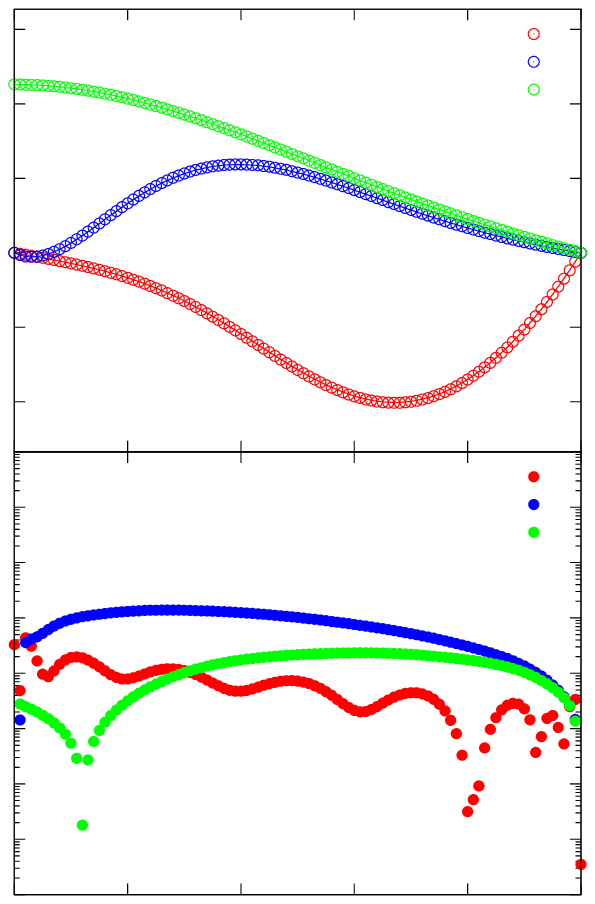}}
\resizebox{7cm}{!}{\include{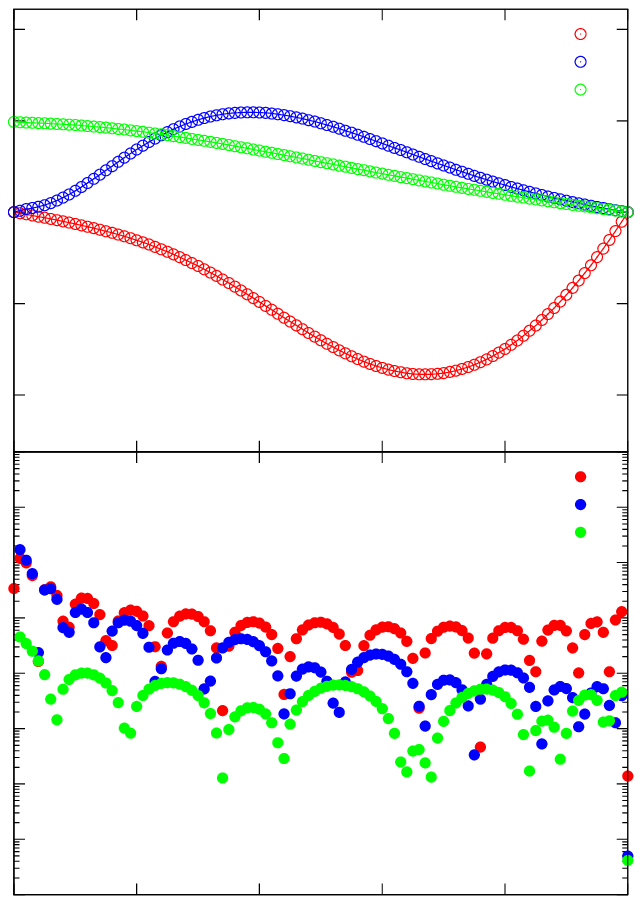}}
\caption{\label{fig:Linstack} (Color Online) 
Comparison of the fitted models to the linear coupling solution (top) and the exponential coupling solutions (bottom) for $\alpha = 0.005$ (left) and $\alpha = 0.010$ (right). Included are the rescaled numerical solution and fits (top) and the resulting fit residual (bottom). Color indicates the field component and the solid line indicates the analytical fitted model, whereas dots indicate numerical data points. The field components have been propelry rescaled, as indicated in the legend. Note that for both coupling strengths the model fit residual is below our desired tolerance throughout the entire $x$ domain.}
\end{center}
\end{figure*}

A comparison between the numerical data and the analytical fitted models is presented in Fig.~\ref{fig:Linstack}. Here we plot the field components (properly rescaled to fit all in the same figure) in each top plot and their corresponding residuals as a function of the compactified coordinate $x$ in each bottom plot for both the linear sGB and EdGB solutions and for two coupling values of $\bar{\alpha}$. We find that the residual between the numerical data and the analytical fitted model is always below our desired tolerance on the numerical solution of $\mcl{O}(10^{-5})$. With this caveat, the fitted models can be treated as ``exact" for practical applications (up to the accuracy mentioned above).
%

\subsection{ISCO}

The inner edge of accretion disks around black holes are typically characterized by the innermost stable circular orbit (ISCO) of massive test-particles~\cite{Abramowicz2013}. For our spherically symmetric ansatz, the marginal stable circular orbits are determined by solving for circular timelike geodesics for massive test-particles~\cite{PhysRevD.83.124043, PhysRevD.90.084009} orbiting the black hole. This is equivalent to requiring conditions on the effective potential, $\dot{r}^2 = V_{\eff}$, where this potential is given by
\be
V_{\eff} = \frac{1}{m(\rho)} \lb -1 - \frac{L^2}{m(\rho) \rho^2} + \frac{E^2}{f(\rho)} \rb,
\ee
where $E$ and $L$ are the energy and angular momentum per unit mass (for massive particles) respectively, defined from the conserved quantities corresponding to the temporal and azimuthal Killing vectors in a stationary and spherically symmetric spacetime.

As in Newtonian gravity, a stable or an unstable circular orbit occurs at local minima or maxima of the effective potential, such that $\dot{r} = V_{\eff} = 0$ and $V_{\eff}^{\prime} = 0$. In a Schwarzschild metric, there is both an unstable and a stable circular orbit, such that the unstable orbit is closer to the horizon, and the distance between these orbits is determined by the angular momentum $L$. The innermost stable circular orbit is equivalent to finding the value of the angular momentum where these stable and unstable orbits coincide (because the unstable orbit will always be closer to the horizon than the stable orbit). By analogy, requiring these orbits coincide is equivalent to finding the saddle points of the effective potential, which are located at $V_{\eff}^{\prime \prime} = 0$. By combining these three conditions, we find the generalized equation
\be
\label{eq:ISCOcond}
\lb \frac{1}{f (\rho)} \rb^{\prime} \lb \frac{1}{\rho^2 m(\rho)} \rb^{\prime \prime} - \lb \frac{1}{\rho^2 m(\rho) } \rb^{\prime} \lb \frac{1}{f(\rho)} \rb^{\prime \prime} = 0,
\ee
the solutions (there may be multiple) of which give the locations of the marginal stable circular orbits of the spacetime. The smallest of these solutions is identified as the ISCO. 

In our compactified isotropic coordinates, the location of the ISCO in GR is $x_{\ISCO}^{\GR} = \sqrt{6}/3$ which corresponds to the familiar $r_{\ISCO}^{\GR} = 3 \, r_{\Hz} = 6 \, M_{0}$ when transformed to Schwarzschild coordinates. In sGB gravity, the ISCO location is shifted from this Schwarzschild value. We can find the ISCO shift using the perturbative solution of Eq.~\eqref{eq:linsolx} to find
\be
\label{eq:isco-pert}
x_{\ISCO} = x_{\ISCO}^{\GR} \lsb 1 + \frac{\bar{\alpha}^2}{\beta}  \lb \frac{427634}{841995} + \frac{2383}{13860} \sqrt{6} \rb \rsb,
\ee
which is identical to that of~\cite{PhysRevD.83.104002} when converted to Schwarzschild coordinates. We can also find the ISCO shift for the numerical metric solving Eq.~\eqref{eq:ISCOcond} with a Newton-Raphson algorithm. By taking the location of the ISCO in GR as our initial guess, we ensure that the converged root is the desired root, as we expect deviations to be comparably small. 

We presented these results already on the left panel of Fig.~\ref{fig:ISCOLRcomp} in Sec.~\ref{sec:intro}, where we saw that the ISCO shift is typically smaller than $10^{-4}$. We also saw there that the shift computed with the analytic perturbative solution in the linear sGB case [Eq.~\eqref{eq:isco-pert}] agrees well the shift computed with the numerical solution in linear sGB but disagrees in EdGB. Interestingly, the shift computed with the fitted models agree extremely well with the numerical solution in both cases.  

\subsection{Light Ring}

The light ring or photon sphere is the surface generated by all unstable circular null geodesics of photons. The location of the light ring around black holes is important for observations with the Event Horizon Telescope~\cite{2009astro2010S..68D}, which is imaging the black hole shadow of Sagittarius A*, i.e.~the electromagnetically dark region caused by photons that cross the light ring and fall into the event horizon. Future observations of black hole shadows may be able to place constraints on the location of the light ring in other quadratic gravity theories~\cite{0264-9381-35-23-235002}. 

Similar to the ISCO calculation, the light ring can be found by requiring certain conditions on the effective potential. For massless particles, there is only a single unstable circular orbit and no stable circular orbits. Thus to find the unstable circular orbit, we need only require $V_{\eff} = 0$ and $V_{\eff}^{\prime} = 0$, which leads to the equation
\be
\label{eq:LRcond}
f(\rho) \lb \frac{1}{f(\rho)} \rb^{\prime} - \rho^2 m(\rho) \lb \frac{1}{\rho^2 m(\rho)} \rb^{\prime} = 0.
\ee
As before, the smallest solution to this equation returns the location of the light ring around the black hole. 

The location of the light ring in GR is simply $x_{\LR}^{\GR} = \sqrt{3}/3$, which reduces to $r_{\LR} = 1.5 \, r_{\Hz} = 3 \, M_{0}$ in Schwarzschild coordinates. As in the ISCO case, the location of the light ring is shifted in sGB gravity. We can calculate this shift with the perturbative analytic solution to find 
\be
x_{\LR} = x_{\LR}^{\GR} \lsb 1 + \frac{\bar{\alpha}^2}{\beta}  \lb -\frac{189328}{841995} + \frac{2383}{3465} \sqrt{3} \rb \rsb.
\ee
a result that to the best of our knowledge had not appeared in the literature previously. In the full non-linear case, we must solve Eq.~\eqref{eq:LRcond} numerically using a Newton-Raphson method with the GR shift as our initial guess. 

The light-ring shift was already presented on the right panel of Fig.~\ref{fig:ISCOLRcomp}. The shift is comparable to the shift of the ISCO, typically smaller than $10^{-4}$. Interestingly, we do find a noticeable disagreement between the analytic perturbative solution and the linear coupling case for higher values of $\bar{\alpha}$, which was not present in the other calculated observables. The comparison between these two solutions in Fig.~\ref{fig:LinCgtt} shows that the largest differences between them occur closer to the horizon. Therefore this difference is larger in the region around the location of the light ring ($x_{\LR}^{\GR} \approx 0.57$) than in both the region around the location of the ISCO ($x_{\ISCO}^{\GR} \approx 0.82$) and asymptotically far away ($x \approx 1$). Thus it is expected that an observable calculated in this region should have a comparatively magnified discrepancy between the analytic perturbative solution and the numerical linear sGB solution. We also find that the fitted models agree extremely well with the numerical solutions for both coupling cases.

\subsection{Naked Singularity}
\label{ssec:naked}

\begin{figure*}[htb]
\begin{center}
\resizebox{8cm}{!}{\include{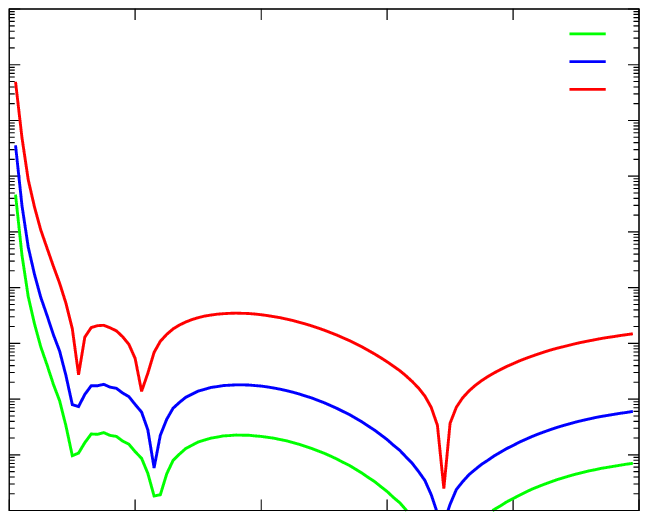}}
\resizebox{8cm}{!}{\include{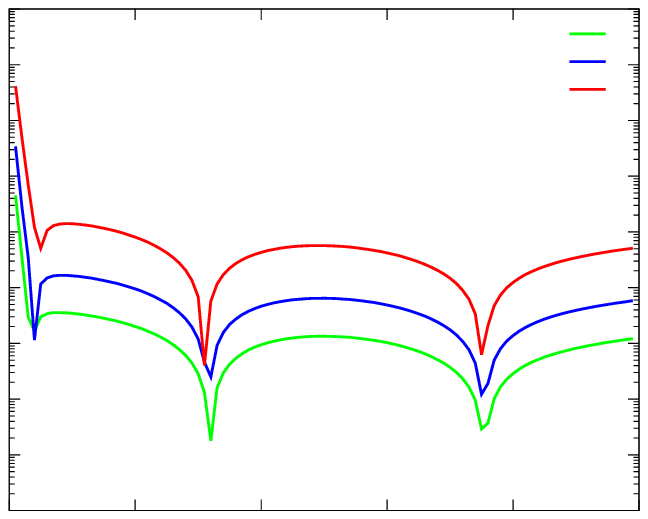}}
\caption{\label{fig:LinCGBdiff} Logarithm of normalized difference between the Gauss-Bonnet curvature invariant computed with the analytical perturbative (superscript ``P") solution and the numerical solution in the linear coupling theory (left panel) and the exponential coupling theory (right panel).}
\end{center}
\end{figure*}

Spherically symmetric black holes in sGB gravity have been shown to possess a minimum size for a given $\bar{\alpha}$ in both EdGB~\cite{PhysRevD.54.5049} and linear sGB~\cite{PhysRevD.90.124063}. This results from a consistency condition on the field equations, obtained by requiring the scalar field to be regular on the horizon. Physically, as one increases the coupling strength $\bar{\alpha}$, the location of the curvature singularity inside the horizon grows while the location of the event horizon shrinks, until at some critical value of $\bar{\alpha}$ the two coincide. For values of $\bar{\alpha}$ larger than this critical value, the curvature singularity is outside the event horizon, leading to a naked singularity. Requiring that the latter do not exist yields a maximum value of the coupling strength (and a minimum size of the event horizon) for which sGB black hole solutions can exist. 

Our numerical solutions confirm these results. In our numerical calculations, we impose boundary conditions on the horizon using compactified coordinates at $x = 0$. The transformation from Schwarzschild to compactified coordinates absorbs the horizon shift, so that physical the horizon is always located at $x =0$ in our numerical grid. This then implies that there is a maximum value of $\bar{\alpha}$ above which black hole solutions should not exist in our numerical code. Indeed, we find that for values of $\bar{\alpha}$ larger than roughly $\bar{\alpha}^{\, \star} \approx 0.0131$ on our grid of $N = 101$ points, our code ceases to converge to the required tolerance. This is because a curvature singularity sufficiently near (or inside) the computational domain induces large errors in the Newton polynomial representation of the solution near the horizon boundary, which then propagates through the entire domain in each iteration, preventing the algorithm from converging. 

This is indeed what we see in our numerical calculations as we increase $\bar{\alpha}$: the estimated discretization error on the horizon ($x=0$) begins to grow as the location of the curvature singularity approaches the event horizon boundary. Eventually, the curvature singularity is close enough to the horizon radius that the discretization error near the horizon becomes too large for the specified tolerance. In order to ensure that these results are not a numerical artifact, we implemented adaptive step size refinement on the computational grid that is triggered if the discretization error becomes too large. Even with this adaptive measure in place, the discretization error still grows near the horizon for sufficiently large $\bar{\alpha}$, preventing the code from converging.   

In  order to further support these conclusions, we have computed the Gauss-Bonnet curvature invariant for different values of $\bar{\alpha}$ in both the linear sGB and EdGB theories. Figure~\ref{fig:LinCGBdiff} shows this invariant as a function of the compactified coordinate $x$. Observe that as $x \to 0$ (near the horizon) the curvature invariant begins to grow to ever larger values as $\bar{\alpha}$ is increased.  Observe that as $\bar{\alpha}$ increases, the correction to $\mcl{G}$ at orders larger than $\mcl{O}(\alpha^2)$ begin to quickly approach the correction at $\mcl{O}(\alpha^2)$ where this effect is not present.

\section{Conclusions}
\label{sec:concs}

We have here developed a new numerical framework to solve for stationary and spherically-symmetric spacetimes that represent black holes in a wide class of modified theories of gravity. This framework uses a Newton polynomial representation for the discretized functions, it then recasts the differential system as a linear algebra problem, and then solves the latter through a relaxed Newton-Raphson iterative method. Through the successive minimization of the residual, our framework is capable of controlling the maximum error in the final numerical solution. We have validated this framework through a toy problem consisting of a simple differential equation, through the Schwarzschild metric and by investigating black holes in sGB gravity.

With the sGB solutions at hand, we then investigated a series of physical properties of these spacetimes. First, we verified that the differences between exact numerical solutions and analytic perturbative solutions are very small when the coupling is linear. We then also verified that the exact numerical solutions in linear sGB differ quite significantly from those in EdGB. These similarities and differences manifest themselves not only through the metric tensor, but also through physical observables like the ADM mass, the scalar charge, the location of the ISCO, and that of the light-ring. We finally verified that sGB black holes do not exist beyond a critical value of the sGB coupling, as beyond this value a naked curvature singularity arises. 

We then concluded our analysis by developing analytic fitting functions for the numerical solutions. These fitting functions are constructed through a combination of controlling factors (inspired by the analytic perturbative solution) and polynomials in the compactified coordinate. We verified that the fitting functions agree with the numerical solutions up to the numerical error in the latter. We then computed physical observables, like the ADM mass, the location of the ISCO and that of the light ring with the fitting functions and found excellent agreement between these results and those obtained from the exact numerical solutions. 

The work we did here now opens the door to several further studies. The fitting functions described above, for example, could be used as the analytical background on which to study polar and axial perturbations. Such perturbations would then reveal the quasi-normal mode spectrum of sGB black holes for arbitrary values of the coupling. The quasi-normal mode frequencies could then be used to carry out spectroscopic tests of General Relativity with gravitational wave observations of merging black holes (provided the merger remnant has very small final spin).  

Another interesting direction for future research is the extension of the methods developed here to axisymmetric black holes. The computational infrastructure we presented here is easily extendable to this case. The system of equations of course becomes more complicated not only because new metric functions must be solved for, but also because these functions will depend on both radius and polar angle. We have already extended the work presented in this paper to a two-dimensional grid that is capable of solving for the Kerr metric in General Relativity, and thus, we expect that extensions to modified gravity at this point should be straightforward. 

Once such solutions are found, the fitting methodology developed here could be implemented in the axisymmetric case to find fully analytic approximations for all components of the metric tensor. Such a solution could then be used once more as a background on which to study the evolution of perturbations. The quasi-normal spectrum of these perturbations could then be used to place constraints on a variety of modified gravity theories through the future observations of gravitational wave ringdown modes with advanced detectors.

\acknowledgments

We would like to acknowledge Hector Okada-Da Silva for useful comments and suggestions. A.~S.~ and N.~Y.~would like to acknowledge support from the NSF CAREER grants PHY-1250636 and PHY-1759615, as well as NASA grants NNX16AB98G and 80NSSC17M0041. T.~P.~S.~acknowledges partial support from the STFC Consolidated Grant No.~ST/P000703/1 and  networking support from the COST Action GWverse CA16104.

\appendix
\section{}
\label{app:coeff}

Here are a few fitting coefficients of Eq.~\eqref{eq:nonfit}. The full tables are available by request in {\texttt{Mathematica}}.

\begin{table}[!htb]
\caption{Fitting Coefficients for $a_{i,j}$} 
\begin{tabular}{c c c c} 
\hline\hline 
$i$ & $j$ & sGB & EdGB \\ [0.5ex] 
\hline 
0 & 2 & \num{0.0} & \num{-4.08203E-4} \\ 
1 & 2 & \num{0.0} & \num{5.97831E-3} \\
3 & 2 & \num{0.0} & \num{-2.84305E-2} \\
\vdots & \vdots & \vdots & \vdots \\
0 & 3 & \num{0.0} & \num{9.2781E1} \\
\vdots & \vdots & \vdots & \vdots \\
0 & 4 & \num{-2.9618E1} & \num{-7.97313E4}\\
1 & 4 & \num{6.00622E2} & \num{6.27646E5}\\
3 & 4 & \num{6.34509E3} & \num{-1.47352E6}\\
\vdots & \vdots & \vdots & \vdots \\
0 & 6 & \num{1.62176E6} & \num{2.26674E10}\\
\vdots & \vdots & \vdots & \vdots \\
13 & 12 & \num{3.70117E17} & \num{8.55071E21} \\ [1ex]
\hline 
\end{tabular}
\label{table:aij} 
\end{table}

\begin{table}[!htb]
\caption{Fitting Coefficients for $b_{i,j}$} 
\begin{tabular}{c c c c} 
\hline\hline 
$i$ & $j$ & sGB & EdGB \\ [0.5ex] 
\hline 
0 & 2 & \num{0.0} & \num{-3.25825E-1} \\ 
1 & 2 & \num{0.0} & \num{-2.53488E1} \\
\vdots & \vdots & \vdots & \vdots \\
0 & 3 & \num{0.0} & \num{-8.77688E2} \\
\vdots & \vdots & \vdots & \vdots \\
0 & 4 & \num{-396346E3} & \num{-4.45446E5} \\
1 & 4 & \num{2.46921E4} & \num{-3.75943E7} \\
\vdots & \vdots & \vdots & \vdots \\
0 & 6 & \num{1.44775E9} & \num{1.02022E10} \\
\vdots & \vdots & \vdots & \vdots \\
11 & 10 & \num{2.16553E19} & \num{-1.23741E21} \\
\vdots & \vdots & \vdots & \vdots \\
11 & 12 & \num{-1.98032E23} & \num{0.0} \\ [1ex]
\hline 
\end{tabular}
\label{table:bij} 
\end{table}

\FloatBarrier
\begin{table}[!htb]
\caption{Fitting Coefficients for $c_{i,j}$} 
\begin{tabular}{c c c c} 
\hline\hline 
$i$ & $j$ & sGB & EdGB \\ [0.5ex] 
\hline 
0 & 1 & \num{0.0} & \num{-3.19873E-4} \\ 
1 & 1 & \num{0.0} & \num{6.33399E-3} \\
\vdots & \vdots & \vdots & \vdots \\
0 & 2 & \num{0.0} & \num{4.31769E1} \\
\vdots & \vdots & \vdots & \vdots \\
0 & 3 & \num{1.20972E2} & \num{-3.06323E1} \\
1 & 3 & \num{2.92929E1} & \num{1.02085E4} \\
\vdots & \vdots & \vdots & \vdots \\
0 & 5 & \num{6.39937E5} & \num{ -3.76357E7} \\
\vdots & \vdots & \vdots & \vdots \\
12 & 8 & \num{0.0} & \num{3.02982E17} \\
\vdots & \vdots & \vdots & \vdots \\
7 & 11 & \num{-5.00065E18} & \num{0.0} \\ [1ex]
\hline 
\end{tabular}
\label{table:cij} 
\end{table}
%

\bibliographystyle{apsrev}
\bibliography{NumerEDGBBiblio}
\end{document}